\documentclass[sigconf]{acmart}

\usepackage[mathcal]{eucal}
\usepackage{bm}
\usepackage{multirow}
\usepackage{colortbl}       % colored line in table
\usepackage{enumitem}
\usepackage[utf8]{inputenc}

\AtBeginDocument{%
  \providecommand\BibTeX{{%
    \normalfont B\kern-0.5em{\scshape i\kern-0.25em b}\kern-0.8em\TeX}}}

\copyrightyear{2022} 
\acmYear{2022} 
\setcopyright{rightsretained} 
\acmConference[CHI '22]{CHI Conference on Human Factors in Computing Systems}{April 29-May 5, 2022}{New Orleans, LA, USA}
\acmBooktitle{CHI Conference on Human Factors in Computing Systems (CHI '22), April 29-May 5, 2022, New Orleans, LA, USA}
\acmDOI{10.1145/3491102.3501826}
\acmISBN{978-1-4503-9157-3/22/04}

\acmSubmissionID{9955}

\begin{document}

% \title{Explainable Speech Emotion Recognition with Contrastive Rationalization}
% \title{Towards Relatable Contrastive Explainable AI with the Perceptual Process}
\title{Towards Relatable Explainable AI with the Perceptual Process}

\author{Wencan Zhang}
\affiliation{%
  \institution{National University of Singapore}
%   Department of Computer Science, 
  \streetaddress{COM1, 13, Computing Dr}
  \city{}
%   \state{State}
  \country{Singapore}
  \postcode{117417}
}
\email{wencanz@u.nus.edu}

\author{Brian Y. Lim}
\affiliation{%
  \institution{National University of Singapore}
  \streetaddress{COM1, 13, Computing Dr}
  \city{}
%   \state{State}
  \country{Singapore}
  \postcode{117417}
}
\email{brianlim@comp.nus.edu.sg}

% \author[1]{Wencan Zhang}
% \author[2]{Brian Y. Lim}
% \email[1]{wencanz@u.nus.edu} 
% \email[2]{brianlim@comp.nus.edu.sg}
% % \author{Wencan Zhang \\ wencanz@u.nus.edu \and Brian Y. Lim \\ brianlim@comp.nus.edu.sg}
% \affiliation{%
%   \institution{National University of Singapore}
%   \streetaddress{COM1, 13, Computing Dr}
% %   \city{Singapore}
% %   \state{State}
%   \country{Singapore}
%   \postcode{117417}
% }

% \renewcommand{\shortauthors}{First Author’s Name, Initials, and Last Name}

\begin{abstract}
Machine learning models need to provide contrastive explanations, since people often seek to understand why a puzzling prediction occurred instead of some expected outcome. 
Current contrastive explanations are rudimentary comparisons between examples or raw features, which remain difficult to interpret, since they lack semantic meaning. We argue that explanations must be more relatable to other concepts, hypotheticals, and associations. 
Inspired by the perceptual process from cognitive psychology, we propose the XAI Perceptual Processing Framework and RexNet model for relatable explainable AI with Contrastive Saliency, Counterfactual Synthetic, and Contrastive Cues explanations. 
We investigated the application of vocal emotion recognition, and implemented a modular multi-task deep neural network to predict and explain emotions from speech. 
From think-aloud and controlled studies, we found that counterfactual explanations were useful and further enhanced with semantic cues, but not saliency explanations.
This work provides insights into providing and evaluating relatable contrastive explainable AI for perception applications.
\end{abstract}

% \keywords{speech emotion recognition, neural networks, explainable AI}
% \keywords{Explainable AI, contrastive explanations, vocal emotion, evaluations}

\maketitle

% To add: our framework belongs to "anti-hoc" XAI 
% some other aspect to consider: 
% F3 Explanation target: both data and model’s prediction,
% F4 Explanation scope: both local and cohort (contrastive cues reflect the relationship between classes, thus cohort)
% F7 Relation to the predictive system: anti-hoc
% O2 Explanation medium: mixed of visualization, exemplars, and rationalization
% O3 Explanation interaction: Interactive (users can change to different classes for comparison)
% O6 Explanation audience: lay-users

% remember to cite Jeff R's CSCW2021 paper

\section{Introduction}

With the increasing availability of data, deep learning-based artificial intelligence (AI) has achieved strong capabilities in computer vision \cite{krizhevsky2012imagenet}, natural language processing \cite{kenton2019bert}, and speech processing \cite{amodei2016deep}. 
However, their complexity limits their use in real-world applications due to the difficulty to understand them~\cite{gunning2019darpa}.
To address this, much research has been conducted on explainable AI (XAI) to develop new XAI algorithms and techniques~\cite{adadi2018peeking, arrieta2020explainable, guidotti2018survey, hohman2018visual}, understand user needs~\cite{liao2020questioning, lim2009assessing, ehsan2021expanding, miller2019explanation, wang2019designing} and evaluate their helpfulness~\cite{abdul2020cogam, lim2009and, cai2019effects, cai2019human, coppers2018intellingo, ehsan2018rationalization, poursabzi2021manipulating, wang2021show}.

Despite the myriad XAI techniques, many of them remain difficult to understand, due to the lack of human-centered design that do not satisfy human needs~\cite{eiband2018bringing,miller2019explanation,wang2019designing}.
Miller identified contrastive reasoning as a particular reason that people ask for explanations~\cite{miller2019explanation} --- \textit{"one does not explain events per se, but that one explains why the puzzling event occurred in the target cases but not in some counterfactual contrast case.}"~\cite{hilton1990conversational}.
We further argue that explanations lack \textit{relatability} towards concepts that people are familiar with, and therefore they seem too low-level technical and not semantically meaningful.
Existing contrastive explanation techniques~\cite{dhurandhar2018explanations, van2018contrastive, miller2021contrastive, le2020grace} remain unrelatable, hence limiting their interpretability.
In this work, we extend the framing of relatable explanations beyond contrastive explanations to include saliency, counterfactuals, and cues.
Explanations should be relatable towards \textit{concepts} via contrastive explanations, towards \textit{exemplars} by providing counterfactual examples, and towards \textit{associated auxiliary concepts} such as sensory and semantic cues.

We have identified audio prediction as a problem space in dire need of relatable explanations.
% Our second challenge is how to intuitively explain audio predictions. 
Much research on XAI techniques focuses on structured data with semantically meaningful features, unstructured data such as text with semantically meaningful words or sentences, and images with visualizations that are visually intuitive. 
% Indeed, most explanations are visualized\footnote{Even attribution explanations on words in sentences rely on visual highlighting.}.
Explaining audio visually is problematic, since sound is not visual and people understand them through relating to concepts or other audio samples~\cite{pell2011time}.
Current explanation techniques for audio typically present saliency maps on audiograms or spectrograms. Spectrograms are too technical for lay users or even non-engineering domain experts. Saliency maps are too simple and merely point to regions without explaining their relevance.
Example-based explanations extract or produce examples for users to compare, but this still requires people to speculate why some examples are similar or different.
Hence, explaining audio predictions requires relating the prediction to other concepts, counterfactual examples, and associated cues. 
% This is a remarkable use case to showcase our perceptual processing approach to providing the explanation triplet of contrastive saliency, counterfactual synthetic, and contrastive cues.
We study the use case of vocal emotion recognition to propose relatable explanations.
With applications in smart speakers for the home~\cite{maharjan2019hear}, digital assistants for mental health monitoring~\cite{wang2014studentlife, ben2015next}, and affective computing~\cite{picard2000affective}, there is a growing need for these AI models to be relatably explainable.

Furthermore, not only should explanations be semantically meaningful, but the way the explanations are generated or the way the AI "thinks" should be human-like to earn people's trust~\cite{ross2017right}.
We draw inspiration from theories of human cognition to understand why and how people relate concepts, information, and data.
Specifically, we frame relatable explanations with the Perceptual Process~\cite{carterette_perceptual_1978}, where people select, organize, and interpret information to make a decision. Corresponding to these stages, we propose the \textit{XAI Perceptual Processing Framework} with modular explanations for Contrastive Saliency, Cues, and Counterfactual Synthetics with Contrastive Cues, respectively. 
This was implemented as \textit{RexNet (Relatable Explanation Network)}, a deep learning model with modules for each explanation type.
We evaluated the explanations with a modeling study, a qualitative think-aloud study and a quantitative controlled study to investigate their usage and impact on decision performance and trust perceptions. 
We found that RexNet improved prediction performance and explanation faithfulness; participants appreciated the diversity of explanations; and participants benefited from Counterfactual and Cues explanations, but not for Saliency explanations.
In summary, we address the challenge that explanations need to be relatable, and studied this for an audio prediction task (vocal emotion recognition).
\textbf{Our contributions are:}
\begin{enumerate}[leftmargin=1.4em]
    \item XAI Perceptual Processing Framework for relatable explanations inspired from theories in human cognition.
    \item RexNet model with multiple relatable explanation (Contrastive Saliency, Counterfactual Synthetic, Contrastive Cues).
    \item First to provide relatable explanations for audio predictions.
    \item Evaluation of usage and usefulness of relatable explanations.
\end{enumerate}

\section{Related Work}

We introduce various explainable AI techniques, argue how they lack human-centeredness, and describe the background on speech emotion recognition and highlight their lack of explainability.

\subsection{Explainable AI techniques} 
Much research has been done to develop explainable AI (XAI) for improving model transparency and trustworthiness. 
An intuitive approach is to point out which features are most important. Attribution explanations do this by identifying importance using gradients~\cite{sundararajan2017axiomatic}, ablation~\cite{richardson2006beyond}, activations~\cite{selvaraju2017grad}, or decompositions~\cite{bach2015pixel, montavon2017explaining, ribeiro2016should}.
In computer vision, attributions take the form of saliency maps (e.g., \cite{selvaraju2017grad}).
Explaining by referring to key examples is another popular approach.
This includes simply providing arbitrary samples of specific classes, cluster prototypes or criticisms~\cite{kim2016examples}, or influential training set instances~\cite{koh2017understanding}.
However, users typically have expectations and goals when asking for explanations.

Users ask for contrastive explanations when expected outcomes do not happen.
A simple answer would find the attribution differences between the actual (fact) and expected (foil) outcomes~\cite{prabhushankar2020contrastive}
However, this is naive because users are truly asking for what differences in feature values, not attributions, would lead to the alternative outcome. That is a counterfactual explanation.
Furthermore, to anticipate a future outcome or prevent an undesirable one, users could ask for counterfactual explanations.
Indeed, contrastive explanations are often conflated with counterfactual explanations in the research literature.
Such explanations suggest the minimum changes in the current case to achieve the desired outcome~\cite{wachter2017counterfactual}. 
Trained decision structures, such as local foil trees~\cite{van2018contrastive}, Bayesian rule lists~\cite{letham2015interpretable}, or structural causal models~\cite{miller2021contrastive} can also serve as counterfactual explanations.
Though typically explained in terms of feature values~\cite{wachter2017counterfactual, le2020grace, dhurandhar2018explanations} or anchor rules~\cite{ribeiro2018anchors}, techniques have been developed to synthesize counterfactuals of unstructured data (e.g., images~\cite{goyal2019counterfactual} and text~\cite{hendricks2018generating}).
In this work, we employ the synthesis approach to generate counterfactuals of audio data.

There are many explanation types and Lim and Dey have framed them in an intelligibility taxonomy as Why (Attribution), Why Not (Contrastive), and How To (Counterfactual)~\cite{lim2009assessing}.
Many of these XAI techniques have been independently developed or tested, so their usage is disparate. In this work, we unify them in a common framework and integrate them in a single machine learning model.

% - HCI papers? 
% - Carrie Cai IUI 2019: check works that cite her paper
% - Miller's call for contrastive and counterfactual explanations~\cite{miller2019explanation} had oriented and driven much recent developments such model explanations.
% - We also address this need and presented them unified in a single framework of perceptual processing.
% - Say how they are mostly about tabular or image data, but few about audio or time series, to be discussed next.

\subsection{Human-Centered Explainable AI}
Abdul et al. \cite{abdul2018trends} found a large gap between XAI algorithms and human-centered research.
To close this gap, HCI researchers have been active in evaluating the various benefits of XAI or lack thereof,
including understanding and trust~\cite{lim2009and}, uncertainty~\cite{lim2011investigating, wang2021show, yin2019understanding}, cognitive load~\cite{abdul2020cogam}, types of examples~\cite{cai2019effects}, etc.
Studies have sought to determine the "best" explanation type~\cite{lim2009and, tsai2021exploring}, but others have revealed the benefit of reasoning with multiple explanations~\cite{lim2011design,lim2013evaluating, anderson2020mental}.
Hence, we propose a unified framework to provide multiple relatable explanations together. 
We determined our human-centered explanation requirements by studying literature on human cognition, which is epistemologically similar to works grounded in philosophy and psychology~\cite{miller2019explanation, wang2019designing}, and unlike empirical approaches to elicit user requirements~\cite{ehsan2021expanding, liao2020questioning, lim2009assessing}. 
% Langer et al.~\cite{langer2021we} summarize desiderata of various stakeholders that XAI should satisfy.
%
Furthermore, current works focus on explaining higher-level reasoning tasks, but not perception tasks that are commonplace. This has implications on the depth of explanations to provide, which we investigate in this work.

% {\color{orange}
% Comparatively, human-centric explainable AI can support in a wider range of applications where human and AI need to interact with each other, such as AI-assisted decision-making \cite{zhang2020effect}, human-AI co-creation \cite{louie2020novice, oh2018lead}, AI teaching human \cite{mac2018teaching}.
% }

% Prior works (small little studies): Foucus on one specific type of explantion 
% Danding's: XAI reasaoning, \cite{wang2019designing}
% we do a simialr framework but by unifying multiple explanation 

% Model explanations come in various forms, and there has been much interest to evaluate their efficacy.

% Lim and Dey found that Why and Why Not explanations were more effective than How To and What If explanations~\cite{lim2009and}, and subsequently identified varied usage styles for each explanation type~\cite{lim2011design, lim2013evaluating}

% Show or Suppress~\cite{wang2021show}
% COGAM~\cite{abdul2020cogam}
% Intellingo~\cite{coppers2018intellingo}
% Compared two types of examples (contrastive, ...)~\cite{cai}
% Upol rationalization~\cite{ehsan2018rationalization}
% Doshi-Velez and Kim’s (2017)\cite{doshi2017towards}

% Other IUI, TiiS, CHI, DIS, UIST, CSCW, IMWUT papers? 

\subsection{Speech Emotion Recognition}
Deep learning approaches proliferate research on automatic speech emotion recognition (SER). 
Leveraging the intrinsic time-series structure of speech data, recurrent neural network (RNN) models with attention mechanism have been developed to capture transient acoustic features to understand contextual information~\cite{mirsamadi2017automatic}. 
Employing popular techniques from the computer vision domain, audio data can be treated as 1D arrays or converted to a spectrogram as a 2D image. Convolutional neural networks (CNNs) can then extract spatial features from these audiograms or spectrograms \cite{huang2014speech}. 
Current approaches improve performance by combining CNN and RNN \cite{tzirakis2018end, zhao2019speech}, or modeling with multiple modalities~\cite{yoon2018multimodal}.
%, or using semi-supervised learning {\color{red}or unlabeled training data \cite{neumann2019improving}}. 
Our RexNet model starts with a base CNN model to leverage many more XAI techniques available to CNNs than RNNs. Since our approach is modular, it can be generalized to state-of-the-art SER models.

\begin{figure*}[t]
    \centering
    % \hspace*{-0.5cm}
    \includegraphics[width=12.0cm]{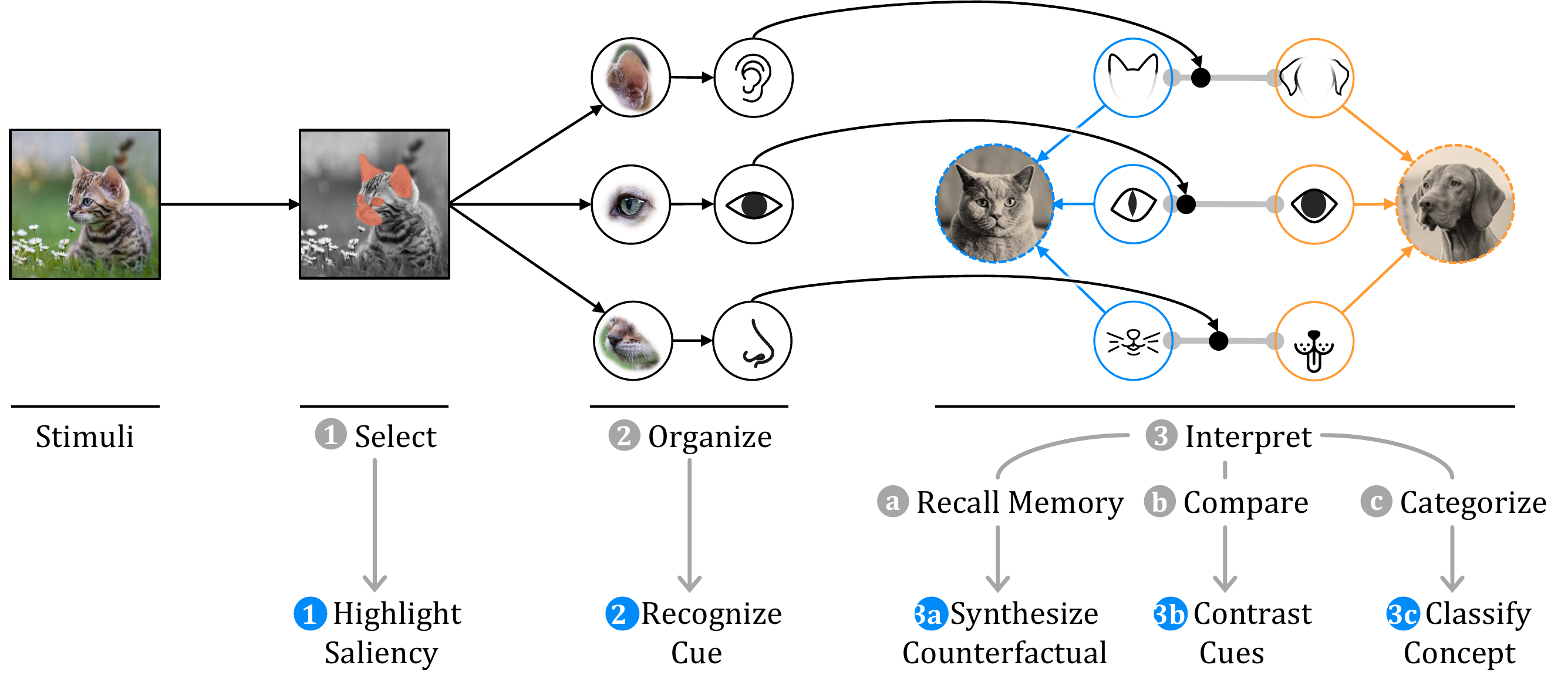}
    \vspace{-0.2cm}
    \caption{
    % Conceptual framework for relatable explainable AI in terms of the human perceptual process.
    XAI Perceptual Processing Framework for relatable explainable AI.
    Inspired by the human perceptual process to select, organize, and interpret stimuli, we propose stages for AI to highlight saliency, recognize cues, and interpret categories (to synthesize counterfactuals, compare cues, classify concepts).
    For visual clarity, we present the use case for visually recognizing a cat instead of a dog, 
    although we use vocal emotion recognition for our prediction task and evaluation.
    Image credits: “dog face” and “cat face” by “irfan al haq”, “Dog” by Maxim Kulikov, "cat mouth" by needumee from the Noun Project.
    }

    \label{fig:perceptualProcessing}
    \Description{The conceptual diagram of XAI Perceptual Processing Framework. The left-most cat photo stands for the visual Stimuli. The next cat photo indicates how people Select the important pixels, which relates to Saliency explanation. Three facial regions with icons (ear, eye, nose) describe how people Organize selected information, which relates to Cue recognition. The last figure with cat, dog and their corresponding icons for ear, eye, nose reveals how people recall familiar animals from memory (relates to Counterfactual explanation), compare the organized information with memory (relates to Contrastive cue explanation), and finally categorize the photo into a concept.}
    \vspace{-0.2cm}
\end{figure*}

\subsection{Model Explanations of Audio Predictions}
Due to the availability of image data and intuitiveness of vision, much XAI research has focused on image prediction tasks; in contrast, few techniques have been developed for audio prediction tasks.
Many techniques exploit CNN explanations by generating a saliency map on the audio spectrogram~\cite{agrawal2020interpretable, krug2018introspection}. Other explanations focus on model debugging by visualizing neuron activations \cite{krug2018neuron}, or as feature visualizing~~\cite{li2020does} (like \cite{olah2017feature} for image kernels).
We also leverage saliency maps as one explanation, due to its intuitive pointing, but augment it with relatable explanations.
Other than explaining the model behavior post-hoc, another approach is to make the model more interpretable and trustworthy by constraining the trained model with domain knowledge, such as with voice-specific parametric convolutional filters \cite{ravanelli2018speaker, loweimi2019learning}.
Our approach with modular explanations of specific types follows a similar objective.

\section{Intuition and Background}

To improve trust, models should provide explanations that are relatable and human-like. Thus, we propose to use theories of human perception and cognition to define our explainable AI techniques. 
We discuss how the framework supports relatable explanations, and
apply it to vocal emotion recognition.
Next, we describe background theories from cognitive psychology and vocal emotion prosody, and define requirements for relatable explanations.

\subsection{Perceptual Processing}

The perceptual process defines three stages for how humans perceive and understand stimuli: selection, organization, and interpretation \cite{carterette_perceptual_1978}. Fig. \ref{fig:perceptualProcessing} illustrates these stages for the case of visually perceiving a cat and relates them to our technical approach.
When sensory stimuli (e.g., light rays or audio vibrations) reach the senses, 1) our brain first \textit{selects} only a subset of the information to focus attention. This is equivalent to highlighting salient regions in an image.
2) The next stage \textit{organizes} the salient regions into meaningful cues. For a face, these would include recognizing the ears, eyes, and nose.
3) Finally, the brain \textit{interprets} these lower-level cues towards higher-level concepts. In our example, the face cues are used to recognize the animal by: a) recalling from long-term memory the concepts of cat and dog, and their respective cues, b) compare whether each element is closer to the cat or dog version (Fig. \ref{fig:perceptualProcessing} uses a slider paradigm for illustration), and c) categorize the concept with the smallest difference.
For our application in vocal emotion recognition, the perceptual processing framework aligns with Kotz et al's model for processing emotion prosody \cite{pell2011time,schirmer2006beyond} that describes stages for "\textit{extracting} sensory/acoustic features, \textit{detecting} meaningful relations, \textit{conceptual processing} of the acoustic patterns in relation to emotion-related knowledge held in long-term memory."

In particular, people categorize concepts by mentally recalling examples and comparing their similarities~\cite{goldstein2014cognitive}. These examples may be prototypes or exemplars. 
With Prototype Theory, people summarize and recall average examples, but these may be quite different from the observed case being compared. 
With Exemplar Theory, people memorize and recall specific examples, but this does not scale with inexperienced cases. Instead, people can \textit{imagine} new cases that they have never experienced~\cite{byrne2007rational}.
Moreover, rather than tacitly comparing some ill-defined difference between the examples, people make comparisons by judging similarities or differences along \textit{dimensions} (cues) \cite{moyer1976mental}. 
Categorization can then be done \textit{systematically} with proposition rules or \textit{intuitively} \cite{kahneman2011thinking}, with either sometimes being more effective~\cite{ma2016trust}.

We apply this framework and propose a unified technical approach with contrastive explanation types to align with each stage of perceptual processing: 1) highlight saliency, 2) recognize cues, 3a) synthesize counterfactual, 3b) compare cues, and 3c) classify concept.
% {\color{blue}In this paper, relatability is defined as explanations' ability to associate with operations in human perceptual process (Fig. \ref{fig:perceptualProcessing}). Specifically, highlighted saliency help people select relevant information and reduce the cognitive load (1); recognizing cues help people organize the selected information (2); synthesized counterfactual examples remind people about the representative instances for a certain concept (3a), showing contrastive cues enhance the comparison on the subtle difference between different concepts (3b).}
We further present cue differences as rules and leverage an embedding for emotions to represent intuition (described later).

% \subsection{Relating Explanations to Concepts, Exemplars, and Cues}
\subsection{Desiderata for Relatable Explanations}

Informed by the Perceptual Process, Prototype and Exemplar Theories, we identified requirements that AI explanations of the prediction of an instance should be made more relatable towards:
\begin{itemize}[leftmargin=1.4em]
    \item \textit{Concepts} by relating the predicted concept to other concepts. 
    Contrastive explanations~\cite{lim2009assessing,miller2019explanation} are thus a key foundation for broader relatable explanations.
    \item \textit{Exemplars} by comparing the factual (actual) instance with counterfactual instances of the other concepts. Concepts are abstract, so providing concrete examples can help people to fixate on details and cues for comparison.
    Counterfactual explanations~\cite{dhurandhar2018explanations,miller2019explanation,wachter2017counterfactual} are a first step in identifying marginally different instances with different prediction outcomes, but do not further relate to why the instances are different.
    \item \textit{Cues} by relating how auxiliary concepts or associated cues are different between the factual and counterfactual instances. 
    For perception tasks, this involves highlighting saliency in sensory cues (stimuli; e.g., eyes of a face). 
    For cognition tasks, this involves articulating differences in semantic cues (e.g., interpreted speech rate from phonemes).
    Attribute value explanations
    % \footnote{Not feature attribution explanations that have been popularized in explainable AI / interpretable machine learning (e.g., \cite{ribeiro2016should}).} 
    are popular in recommender systems~\cite{Pu2007TrustinspiringEI} to describe how two products have similar attributes (e.g., both are red in color), but they are used to explain similarity, rather than contrast.
\end{itemize}
Finally, we propose an integrated architecture, RexNet, which is relatable to human reasoning by mimicking parts of human perceptual processing. Together the individual capabilities and overall architecture can improve trust, understanding, and performance by being more relatable.

\subsection{Vocal Emotion Prosody}

% \begin{table}[h!]
\begin{table}[t]
    \small \setlength{\tabcolsep}{2.25pt}
    \caption{
    Vocal cues for emotion recognition.
    }
    \begin{tabular}{lll}
    \hline
    Vocal Cue \cite{juslin2001impact}   & \begin{tabular}[c]{@{}l@{}}Simple Name \end{tabular}  & Description   \\ \hline
    \begin{tabular}[t]{@{}l@{}}High-Frequency\\ Energy (HF 500)\end{tabular} & \begin{tabular}[t]{@{}l@{}}Shrillness \end{tabular}  & \begin{tabular}[t]{@{}l@{}}Proportion of high-frequency energy\\ (cut-off 500 Hz) in the acoustic\\ spectrum, i.e., how much of the\\ speech is high-pitch. \end{tabular}  \\
    Voice Intensity                                                         & Loudness      & \begin{tabular}[t]{@{}l@{}}Mean of sound amplitude, i.e., how\\ loud the person is speaking. \end{tabular}                                                                                    \\
    Mean ($F_0$)                                                             & Average Pitch & \begin{tabular}[t]{@{}l@{}}Mean of fundamental frequency, i.e.,\\ pitch.\end{tabular}                                                                                                                \\
    SD ($F_0$)                                                               & Pitch Range   & \begin{tabular}[t]{@{}l@{}}Standard deviation of fundamental\\ frequency, i.e., pitch variation.\end{tabular}                                 \\
    
    Speech Rate                                                             & Speaking Rate   & \begin{tabular}[t]{@{}l@{}} How quickly the person is speaking\\(words/second), i.e., $1/t_{total}$.\end{tabular}                                                                                                                                   \\
    Pause Proportion                                                        & \begin{tabular}[t]{@{}l@{}}Proportion\\ of Pauses\end{tabular}      & \begin{tabular}[t]{@{}l@{}} Proportion of pauses in the speech, \\ i.e., $t_{pauses} / t_{total}$ \end{tabular}      
    \\ \hline
    \end{tabular}
    \label{table:cueDefinitions}
\end{table}

People recognize vocal emotions based on various vocal stimulus types and prosodic attributes~\cite{lausen2020emotion}, such as verbal~\cite{juslin2001impact} and non-verbal expressions~\cite{sauter2010perceptual} (e.g., laughs, sobs, screams), and lexical~\cite{mariooryad2014compensating} information.
In this work, we focus on vocal cues (prosody) identified by Juslin et al. (e.g., see Table \ref{table:cueDefinitions}). These cues are about how words are spoken, rather than the words themselves (lexical information).
We leverage people's ability to index vocal emotion categories by the pattern of cues~\cite{juslin2001impact} to identify cue differences between different emotions, which we present in our model explanation.
Although people may be able to perceive various vocal cues, they may be unable to relate to them conceptually (e.g., "formant frequency" is technically complex), therefore, we limit cues to familiar everyday concepts. In our user study, we further verified their understandability in a screening test.
For our prediction application, the \textit{concept} to predict is emotion, \textit{cues} are vocal cues for emotion prosody, \textit{cue differences} support dimensional comparisons, and \textit{saliency} is in terms of phonemes or pauses between them.

\section{Technical Approach}
We propose an interpretable deep neural network to predict vocal emotions and provide relatable explanations. We first describe the base prediction model, then specific explanation modules.

% \subsection{Base Prediction Model for Vocal Emotion Recognition}
\subsection{Base Prediction Model}

\begin{figure*}[ht]
    \centering
    % \hspace*{-0.5cm}
    \includegraphics[width=14.4cm]{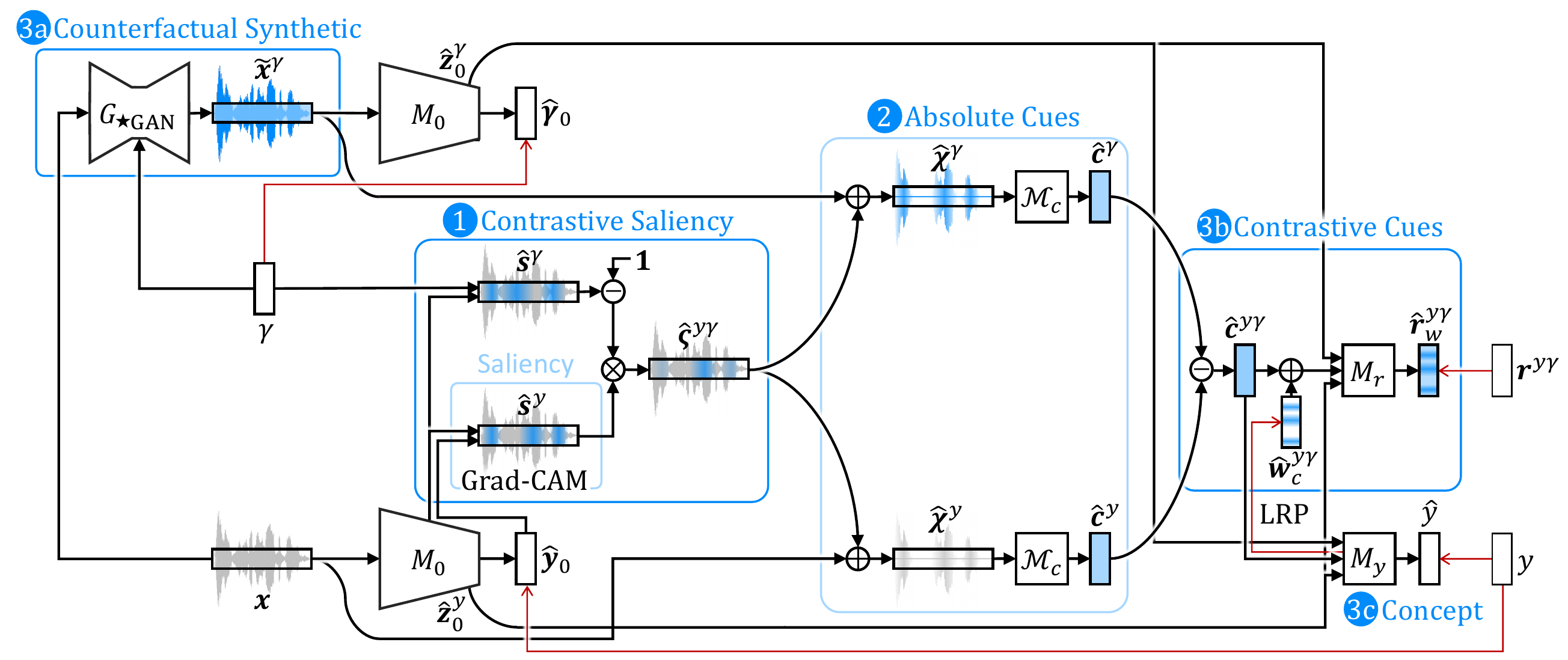}
    \caption{
    Modular architecture of RexNet with relatable explanations for the prediction of emotion $y$ from input voice $\bm{x}$.
    Each module is numbered to match the sequence of the perceptual process (Fig. \ref{fig:perceptualProcessing}).
    Black arrows indicate feedforward activations. Red arrows indicate backpropagation during training.
    The base CNN model $M$ is denoted as a trapezium block to represent its function as an encoder.
    The StarGAN generator $G_{{\star}GAN}$, represented as an encoder-decoder, takes input $\bm{x}$ and output $\tilde{\bm{x}}^\gamma$ with the same shape.
    $\mathcal{M}_c$ is a heuristic model, and $M_r$ and $M_y$ are sub-models with only fully-connected layers.
    % $M_r$ is a relationship threshold estimator which converts numeric difference to categorical relationship (lower, similar, higher).
    Although we trained the model on 2D spectrograms, for illustrative simplicity, the audio data is represented in its 1D audio waveform.
    }
    \label{fig:model}
    \Description{Six components: the backbone module, the Contrastive Saliency module, the Absolute Cues module, the Counterfactual Synthetic module, the Contrastive Cues module, and the Concept classifier. The target audio $\bm{x}$ will firstly pass the backbone module to $M_0$ to obtain the initial prediction $\hat{y}_0$. Meanwhile, the input audio $\bm{x}$ and the counterfactual class label pass the Counterfactual Synthetic module ($G_{{\star}GAN}$) to generate the synthetic audio $\tilde{\bm{x}}^\gamma$. After that, the Contrastive Saliency explanation $\bm{\varsigma}^{y\gamma}$ will be produced by the Contrastive Saliency module, which is later used as a mask to process the extracted Absolute Cue features from the target audio and the synthetic audio. By combining the Absolute Cue features from two audios, the Contrastive Cues module enables predicting the contrastive cue relationship prediction, which in turn are fed as additional information for the Concept classifier.}
    \vspace{-0.3cm}
\end{figure*}

We trained a vocal emotion classifier on the Ryerson Audio-Visual Database of Emotional Speech and Song (RAVDESS) dataset \cite{livingstone2018ryerson} with 7356 audio clips of 24 voice actors (50\% female) reading fixed sentences with 8 emotions (neutral, calm, happy, fearful, surprised, sad, disgust, angry). Each audio clip was 2.5-3.5 seconds long, and we padded or cropped them to a fixed 3.0s.
We parsed each audio file to a time-series array of 48k readings (i.e., 16 kHz sampling rate), and preprocessed it to obtain a mel-frequency spectrogram with 128 frequency bins, 0.04s window size, and 0.01s overlap. 
Treating the spectrogram as a 2D image, we can train a convolutional neural network (CNN) \cite{hershey2017cnn}. Specifically, we trained a CNN with 3 convolutional blocks, and 2 fully connected layers. We used cross-entropy loss for multi-class classification.
In sum, the base CNN model $M_0$ takes audio input $\bm{x}$ to predict an emotion $\hat{y}_0$ (lower left in Fig. \ref{fig:model}).

% \subsection{\color{red}Interpretable Modular Multi-Task Model Architecture}
% \subsection{Relatable Contrastive Explanations}

\subsection{RexNet: Relatable Explanation Network}

\definecolor{color_segments}{HTML}{00B050}
\definecolor{color_category}{HTML}{FF00FF}
\definecolor{color_embedding}{HTML}{CC66FF}
\definecolor{color_cue}{HTML}{ED7D31}
\definecolor{color_counterfactual}{HTML}{4472C4}
\definecolor{color_attribution}{HTML}{C00000}

We introduce RexNet --- \textbf{R}elatable \textbf{Ex}planation \textbf{Net}work\footnote{The 'x' can also stand for a cross for Why Not to indicate contrastive explanations.} --- to provide relatable explanations for contrastive explainable AI.
We extended the base model with multiple modules to provide three relatable contrastive explanations (Fig. \ref{fig:model}). 
% These comprise non-contrastive and contrastive explanations. 
The whole architecture can be understood in terms of a chain of dependencies. We describe this in reverse starting with the goal. Ultimately, we want to explain the prediction with descriptive contrastive cues. This requires a counterfactual "foil" to compare the target "fact" with, therefore, we need to obtain an example for comparison. When making a comparison, not all stimuli are relevant for interpretation, hence, we need to select salient segments. For example, noticing a flower in a photo of a pet is irrelevant to identifying whether an animal is a dog or cat.
% In summary, our interpretable prediction approach has the following steps:
In summary, our approach has steps:
\begin{enumerate}[leftmargin=1.4em]
  \item [1.] Highlight {\color{color_segments}salient segments}
    \begin{enumerate}
      \item [i.] Predict emotion {\color{color_category}concept} as initial estimation
      \item [ii.] Keep {\color{color_embedding}embedding} vector of estimation for final classification
      \item [iii.] Explain \textbf{contrastive saliency} using \textit{discounted} Grad-CAM %\cite{selvaraju2017grad}
    \end{enumerate}
  \item [2.] Describe {\color{color_segments}segments}
    \begin{enumerate}
      \item [i.] Infer associated {\color{color_cue}cues}
    \end{enumerate}
  \item [3a.] Generate {\color{color_counterfactual}counterfactual exemplar} for each contrast {\color{color_category}concept}
    \begin{enumerate}
      \item [i.] Generate \textbf{counterfactual synthetic} using StarGAN-VC \cite{kameoka2018stargan}
    \end{enumerate}
  \item [3b.] Compare {\color{color_cue}cue differences} between target case and each exemplar
    \begin{enumerate}
      \item [i.] Calculate {\color{color_cue}cue differences} weighted by {\color{color_segments}saliency}
      \item [ii.] Classify {\color{color_cue}cue difference relations} with {\color{color_cue}cue differences} and {\color{color_embedding}embedding} for target and contrast concepts.
    \end{enumerate}
  \item [3c.] Classify {\color{color_category}concept} fully
    \begin{enumerate}
      \item [i.] Predict {\color{color_category}concept} using inputs: {\color{color_cue}cue differences} of all {\color{color_counterfactual}counterfactuals} + {\color{color_embedding}embedding} (initial estimation)
      \item [ii.] Explain final {\color{color_category}concept} with {\color{color_attribution}attributions} for {\color{color_cue}cue differences} using Layer-wise Relevance Propagation (LRP)~\cite{bach2015pixel}
    \end{enumerate}
\end{enumerate}
We next describe each module for specific contrastive explanations.

\subsubsection{Contrastive Saliency}

Saliency maps are very popular to explain predictions on images, since they intuitively highlight which pixels the model considers important for the predicted outcome~\cite{simonyan2014deep,selvaraju2017grad,zhou2016learning}. For spectrograms, they can identify which frequencies or time periods are most salient \cite{hershey2017cnn}. 
However, they have limited interpretability, since they merely point to raw pixels but do not further elaborate on why those pixels were important. For time-series data, highlighting on a spectrogram remains uninterpretable to non-technical users, since many are not trained to read spectrograms. Furthermore, some salient pixels may be important across all prediction classes, and thus be less uniquely relevant to the specific class of interest. For example, a saliency map to predict emotions from faces may always highlight the eyes regardless of emotion.
To address the issue of saliency lacking semantic meaningfulness, we introduce \textit{associative cues}, which we describe later. 
Here, we address the need for more specific saliency with a discounted saliency map to produce \textit{contrastive saliency}. This retains some importance of globally important pixels, unlike current methods that simply subtract a saliency map of one class from that of another class \cite{prabhushankar2020contrastive}. Dhurandhar et al. \cite{dhurandhar2018explanations} identified pertinent positives and negatives for more precise contrastive explanations by perturbing features, but our approach calculates based on feature activations.

We define two forms of contrastive saliency: pairwise and total. \textit{Pairwise} contrastive saliency highlights pixels that are important for predicting target class $y$ but discounts pixels that are also important for foil class $\gamma$. We implemented the saliency map with Grad-CAM \cite{selvaraju2017grad}, and define the class activation map for class $y$ as $\mathbf{s}^y$. The pairwise contrastive saliency between classes $y$ and $\gamma$ is thus:
\begin{equation}
    \bm{\varsigma}^{y\gamma} = \bm{\lambda}^{y\gamma} \odot \mathbf{s}^{y}
    \label{eq:pairwiseContrastiveSaliency}
\end{equation}
where $\bm{\lambda}^{y\gamma} = (\mathbf{1} - \mathbf{s}^{\gamma})$ indicates the discount factors for all pixels due to their attributions to class $\gamma$, $\mathbf{1}$ is a matrix of all ones, and $\odot$ is the Hadamard operator for pixel-wise multiplication.
To identify important pixels for class $y$ but not any other class, we define \textit{total} contrastive saliency as:
\begin{equation}
    \bm{\varsigma}^{y} = \bm{\lambda}^{y} \odot \mathbf{s}^{y}
    \label{eq:totalContrastiveSaliency}
\end{equation}
where $\bm{\lambda}^{y} = \sum_{\gamma \in C \setminus {y}}(\mathbf{1} - \mathbf{s}^{\gamma})/{|C-1|}$ indicates the discount factors across all alternative classes, and $C$ is the number of classes.

In RexNet, the saliency explanation is calculated from the initial emotion classifier $M_0$ predicting an initial emotion concept $\hat{y}_0$.
We present contrastive saliency for audio using a 1D saliency bar aligned to words in the speech (see Fig. \ref{fig:ui-screenshot}), which aggregates saliency in the spectrogram across frequencies per time bin. This is more accessible for lay people to understand since it avoids using technical spectrograms or audiograms (audio waveforms).

\subsubsection{Counterfactual Synthetic}

\begin{figure}[t]
    \centering
    % \hspace*{-0.5cm}
    \includegraphics[width=7.4cm]{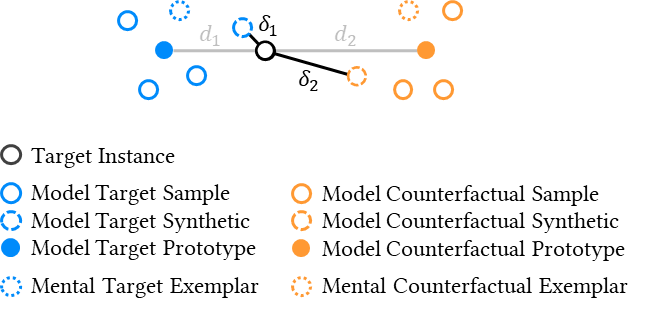}
    \vspace{-0.3cm}
    \caption{
    Conceptual illustration of the benefit of using counterfactual synthetics for comparison.
    Different example types 
    % (sample, synthetic, prototype) 
    have varying distances from the target instance.
    }
    \label{fig:counterfactualEffect}
    \Description{Target Instance refers to the data point to be judged. Model Target / Counterfactual Sample stands for the real data point under the corresponding class. Model Target / Counterfactual Synthetic stands for the synthesized data point. Model Target / Counterfactual Prototype is a virtual data point, which located at the centroid of the class. Mental Target / Counterfactual Exemplar stands for the representative data point in users' mental model.}
    \vspace{-0.3cm}
\end{figure}

Due to the open-ended variability in unstructured data, counterfactual \textit{samples} drawn from a training set are likely to be quite different from the target instance. Counterfactual samples will have extraneous differences that may be distracting to interpret and less meaningful for comparison. Instead, counterfactual \textit{synthetics} are generated to be similar to the target instance, except sufficient differences to achieve the contrastive outcome.
Fig. \ref{fig:counterfactualEffect} illustrates the benefit of using counterfactual synthetics for comparison.
When deciding whether a target item is more similar to a first or second reference, one would measure the target's distance to each reference.
Counterfactual synthesis produces comparison references that are closer to the target item being classified, because it minimizes the differences between the target item and reference example.
These counterfactual synthetics will be closer to other model samples that the model knows (prior instances in the training set), model prototypes (centroids or medoids of class clusters), or human mental exemplars (from the user's memory), since the model may not have a similar example or the human may never have seen or heard a very similar case to the target item.
This amplifies the ratio between the reference distances larger, and makes the difference more perceptible.
Formally, the ratio of differences for counterfactual synthetics are larger than for other examples (prototypes, or samples of prior items), i.e., $|log(\delta_1/\delta_2)| > |log(d_1/d_2)|$.
Therefore, counterfactual synthetics help make comparison between references more easy.

We aim to create a counterfactual that is similar to the target instance $\bm{x}$ which is classified as class $y$, but with sufficient differences to be classified as another class $\gamma$. 
Current counterfactual methods focus on structured (tabular) data by minimizing changes to the target instance~\cite{mothilal2020dice,wachter2017counterfactual}, or identifying anchor rules~\cite{ribeiro2018anchors}, but this is not possible for unstructured data (e.g., images, sounds). 
Instead, inspired by data synthesis with Generative Adversarial Networks (GANs)~\cite{karras2019style,radford2015unsupervised} and style transfer~\cite{choi2018stargan, zhu2017unpaired}, we propose explanations with \textit{counterfactual synthetics} by "re-styling" the original target instance $\bm{x}$ such that it is classified as another class $\gamma$.

\begin{figure}[t]
    \centering
    % \hspace*{-0.5cm}
    \includegraphics[width=5.2cm]{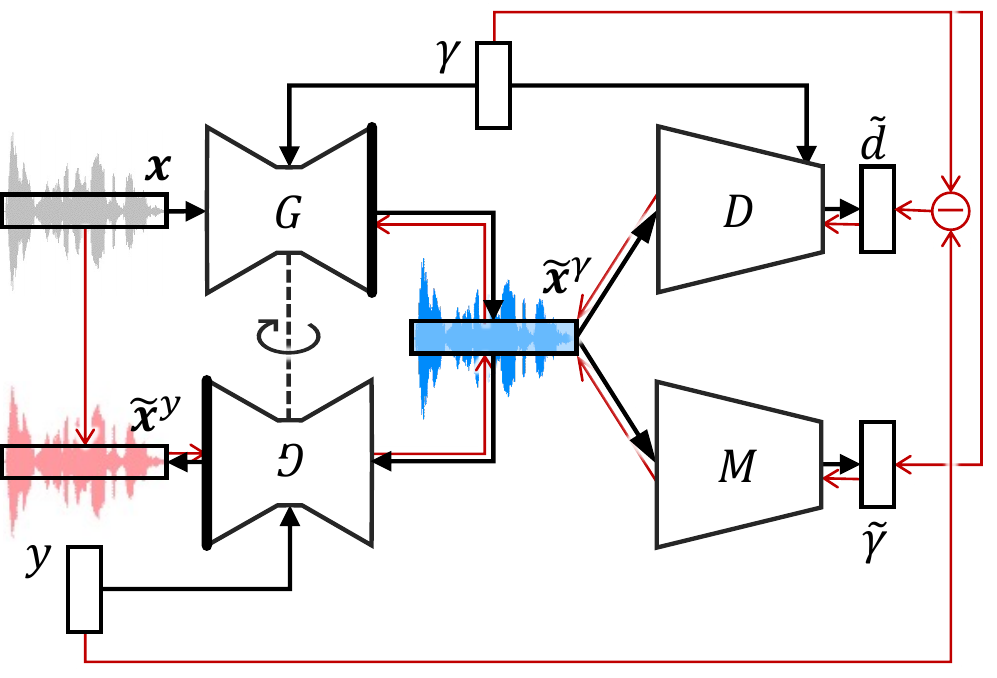}
    \vspace{-0.2cm}
    \caption{
    StarGAN-VC \cite{choi2018stargan} architecture to generate counterfactual synthetics. 
    Black arrows indicate feedforward activations. Red arrows indicate backpropagation during training.
    }
    \label{fig:stargan}
    \Description{Three modules in StarGAN: the generator $G$ produces the synthesized data, the discriminator $D$ judges the quality of synthesized data, the domain classifier $M$ ensures the synthesized data belongs to a certain class. In training, the target audio $\bm{x}$ passes the generator to generate the synthesized audio $\tilde{\bm{x}}^\gamma$, which passes the generator once more to ensure the cycle consistency with original the target audio.}
    \vspace{-0.3cm}
\end{figure}

For vocal emotion recognition, we aim to change the emotion of the speech audio while retaining the original words and identity
% Models to transform instances from one style or class to another have been proposed for binary classification (CycleGAN~\cite{zhu2017unpaired}) 
by using an extension of StarGAN \cite{choi2018stargan} for voice data, StarGAN-VC \cite{kameoka2018stargan}
%, we synthesize a counterfactual instance that is similar to the original instance, but with a different class 
(Fig. \ref{fig:stargan}).
As a generative adversarial model, StarGAN trains three models --- a generator $G$, discriminator $D$, and domain classifier $M$. $G$ inputs the target instance $\bm{x}$ that is of class $y$ and the objective class $\gamma$ to generate a similar instance $\bm{x}^\gamma$. The training objectives are to make $\tilde{\bm{x}}^\gamma \approx \bm{x}$ and $M(\bm{x}^\gamma) \approx \gamma$. 
Next, $\tilde{\bm{x}}^\gamma$ and $y$ are input into $G$ to get $\tilde{\bm{x}}^y$ as output. $G$ is trained to minimize the cycle consistency reconstruction loss between $\bm{x}^y$ and $\bm{x}$, which also improves $\bm{x}^\gamma$. 
$\bm{x}^\gamma$ is also input into the $M$ to output class $\hat{\gamma}$, which is trained to minimize the loss between $\hat{\gamma}$ and $\gamma$. Finally, $D$ is trained to ensure that the generated instances are more realistic $\tilde{d}$.
Together, this semi-supervised method trains $G$ to generate style-transferred instances.

\subsubsection{Contrastive Cues}

The final contrastive explanation involves first inferring cues from the target and counterfactual instances and comparing them. We define the individual cue as \textit{absolute cues} ($\hat{\bm{c}}^y$ and $\hat{\bm{c}}^\gamma$), and the difference as \textit{contrastive cues} $\hat{\bm{c}}^{y\gamma}$.
We report on 6 vocal cues identified by Juslin and Laukka \cite{juslin2001impact} for vocal emotions (Table \ref{table:cueDefinitions}).
Absolute cues can be inferred with machine learning predictions or heuristically.
For vocal emotions, since cues can be deterministically measured from the input data, we use heuristic methods $\mathcal{M_{\mathbf{c}}}$ to infer the cues $\mathbf{c}$. For example, pitch range is calculated as follows: a) calculate fundamental frequency (modal frequency bin) for each CAM-salient
% \footnote{Saliency determined from CAM, above some pre-determined threshold.} 
time window in the spectrogram, b) calculate their standard deviation. For semantically abstract cues, such as sounding "melodic", "questioning", or "nasally", they should be annotated by humans and inferred using supervised learning.

\begin{table}[t!]
    \caption{
    Vocal cues for emotions relative to average levels.
    }
    \vspace{-0.2cm}
    \footnotesize \setlength{\tabcolsep}{1.65pt}
    \begin{tabular}{rllllll}
    \hline
                                                                                    & \multicolumn{6}{c}{Vocal Cue}    \\ \cline{2-7} 
    \multirow{-2}{*}{\begin{tabular}[c]{@{}r@{}}Target \\ Emotion\end{tabular}}      & Shrillness                     & Loudness                       &  \begin{tabular}[c]{@{}l@{}}Average \\ Pitch\end{tabular} & \begin{tabular}[c]{@{}l@{}}Pitch \\ Range\end{tabular} & \begin{tabular}[c]{@{}l@{}}Speaking \\ Rate\end{tabular} & 
    \begin{tabular}[c]{@{}l@{}}Proportion \\ of Pauses\end{tabular} \\ \hline
    
    Neutral                     & {\color[HTML]{D1D1D1} Average} & {\color[HTML]{B8D0EE} Low}     & {\color[HTML]{B8D0EE} Low}                               & {\color[HTML]{B8D0EE} Low}                             & {\color[HTML]{4E679E} High}                              & {\color[HTML]{B8D0EE} Low}                                      \\
    Calm                        & {\color[HTML]{B8D0EE} Low}     & {\color[HTML]{B8D0EE} Low}     & {\color[HTML]{B8D0EE} Low}                               & {\color[HTML]{B8D0EE} Low}                             & {\color[HTML]{D1D1D1} Average}                           & {\color[HTML]{D1D1D1} Average}                                  \\
    Happy                       & {\color[HTML]{4E679E} High}    & {\color[HTML]{4E679E} High}    & {\color[HTML]{4E679E} High}                              & {\color[HTML]{D1D1D1} Average}                         & {\color[HTML]{D1D1D1} Average}                           & {\color[HTML]{B8D0EE} Low}                                      \\
    Fearful                     & {\color[HTML]{D1D1D1} Average} & {\color[HTML]{4E679E} High}    & {\color[HTML]{4E679E} High}                              & {\color[HTML]{4E679E} High}                            & {\color[HTML]{4E679E} High}                              & {\color[HTML]{D1D1D1} Average}                                  \\
    Surprised                   & {\color[HTML]{D1D1D1} Average} & {\color[HTML]{D1D1D1} Average} & {\color[HTML]{D1D1D1} Average}                           & {\color[HTML]{4E679E} High}                            & {\color[HTML]{4E679E} High}                              & {\color[HTML]{B8D0EE} Low}                                      \\
    Sad                         & {\color[HTML]{B8D0EE} Low}     & {\color[HTML]{B8D0EE} Low}     & {\color[HTML]{D1D1D1} Average}                           & {\color[HTML]{D1D1D1} Average}                         & {\color[HTML]{D1D1D1} Average}                           & {\color[HTML]{4E679E} High}                                     \\
    Disgust                     & {\color[HTML]{D1D1D1} Average} & {\color[HTML]{D1D1D1} Average} & {\color[HTML]{B8D0EE} Low}                               & {\color[HTML]{B8D0EE} Low}                             & {\color[HTML]{B8D0EE} Low}                               & {\color[HTML]{4E679E} High}                                     \\
    Angry                       & {\color[HTML]{4E679E} High}    & {\color[HTML]{4E679E} High}    & {\color[HTML]{4E679E} High}                              & {\color[HTML]{4E679E} High}                            & {\color[HTML]{B8D0EE} Low}                               & {\color[HTML]{D1D1D1} Average}                                  \\ \hline
    \end{tabular}
    \label{table:cueDifferences}
    \vspace{-0.2cm}
\end{table}

\begin{table}[t!]
    \caption{
    Contrastive vocal cues for target emotions compared to another emotion (Happy).
    }
    \vspace{-0.2cm}
    \footnotesize \setlength{\tabcolsep}{1.65pt}
    \begin{tabular}{rlllllll}
    \hline
                                & \multicolumn{6}{c}{Vocal Cue}                                                                                                                                                                                                                                                                                    &                                                               \\ \cline{2-7}
    \multirow{-2}{*}{\begin{tabular}[c]{@{}r@{}}Target \\ Emotion\end{tabular}}  & Shrillness                     & Loudness                       & \begin{tabular}[c]{@{}l@{}}Average \\ Pitch\end{tabular} & \begin{tabular}[c]{@{}l@{}}Pitch \\ Range\end{tabular} & \begin{tabular}[c]{@{}l@{}}Speaking \\ Rate\end{tabular} & \begin{tabular}[c]{@{}l@{}}Proportion \\ of Pauses\end{tabular} & \multirow{-2}{*}{\begin{tabular}[c]{@{}l@{}}Contrast\\ Emotion\end{tabular}}  \\ \hline
    Neutral                     & {\color[HTML]{B8D0EE} Lower}   & {\color[HTML]{B8D0EE} Lower}   & {\color[HTML]{B8D0EE} Lower}                             & {\color[HTML]{B8D0EE} Lower}                           & {\color[HTML]{D1D1D1} Similar}                           & {\color[HTML]{D1D1D1} Similar}                                  & Happy                                                         \\
    Calm                        & {\color[HTML]{B8D0EE} Lower}   & {\color[HTML]{B8D0EE} Lower}   & {\color[HTML]{B8D0EE} Lower}                             & {\color[HTML]{B8D0EE} Lower}                           & {\color[HTML]{B8D0EE} Lower}                             & {\color[HTML]{D1D1D1} Similar}                                  & Happy                                                         \\
    Happy                       & {\color[HTML]{D1D1D1} Similar} & {\color[HTML]{D1D1D1} Similar} & {\color[HTML]{D1D1D1} Similar}                           & {\color[HTML]{D1D1D1} Similar}                         & {\color[HTML]{D1D1D1} Similar}                           & {\color[HTML]{D1D1D1} Similar}                                  & Happy                                                         \\
    Fearful                     & {\color[HTML]{D1D1D1} Similar} & {\color[HTML]{D1D1D1} Similar} & {\color[HTML]{D1D1D1} Similar}                           & {\color[HTML]{D1D1D1} Similar}                         & {\color[HTML]{D1D1D1} Similar}                           & {\color[HTML]{4E679E} Higher}                                   & Happy                                                         \\
    Surprised                   & {\color[HTML]{B8D0EE} Lower}   & {\color[HTML]{D1D1D1} Similar} & {\color[HTML]{D1D1D1} Similar}                           & {\color[HTML]{D1D1D1} Similar}                         & {\color[HTML]{D1D1D1} Similar}                           & {\color[HTML]{D1D1D1} Similar}                                  & Happy                                                         \\
    Sad                         & {\color[HTML]{B8D0EE} Lower}   & {\color[HTML]{B8D0EE} Lower}   & {\color[HTML]{B8D0EE} Lower}                             & {\color[HTML]{B8D0EE} Lower}                           & {\color[HTML]{D1D1D1} Similar}                           & {\color[HTML]{4E679E} Higher}                                   & Happy                                                         \\
    Disgust                     & {\color[HTML]{D1D1D1} Average} & {\color[HTML]{B8D0EE} Lower}   & {\color[HTML]{B8D0EE} Lower}                             & {\color[HTML]{B8D0EE} Lower}                           & {\color[HTML]{B8D0EE} Lower}                             & {\color[HTML]{4E679E} Higher}                                   & Happy                                                         \\
    Angry                       & {\color[HTML]{4E679E} Higher}  & {\color[HTML]{4E679E} Higher}  & {\color[HTML]{D1D1D1} Similar}                           & {\color[HTML]{D1D1D1} Similar}                         & {\color[HTML]{B8D0EE} Lower}                             & {\color[HTML]{4E679E} Higher}                                   & Happy                                                         \\ \hline
    \end{tabular}
    \label{table:cueDifferences_vs_happy}
    \vspace{-0.3cm}
\end{table}

We calculated contrastive cues as ordinal cue difference relations $\hat{\bm{r}}_w^{y\gamma}$ from numeric cue differences $\hat{\bm{c}}^{y\gamma}$ based on the instances in the RAVDESS dataset~\cite{livingstone2018ryerson}.
To determine differences between emotions for each cue, we fit the data to a linear mixed effects model with emotion as the main fixed effect and voice actors as random effect (see Supplementary Fig. \ref{fig:cue-distros}), and performed a Tukey HSD test with significance level $\alpha=.005$ to account for the multiple comparison effect. For each cue, if an emotion is not significantly higher than the other, then we label the cue difference as "similar"; otherwise, we label it as "higher" or "lower" depending on the direction.
Table \ref{table:cueDifferences} describes the vocal cue patterns of each emotion compared to average levels, which is in close agreement with \cite{juslin2001impact} except for the fearful emotion.
Table \ref{table:cueDifferences_vs_happy} describes the \textit{pairwise} cue difference relations between each emotion and an emotion (happy).

Predicting the cue difference relations $\hat{\bm{r}}_w^{y\gamma}$ requires deciding the decision threshold at which to split the cue difference $\hat{\bm{c}}^{y\gamma}$ to categorize the relation, and this can contextually depend on initially estimating which emotion concepts $\hat{y}_0$ and $\hat{\gamma}_0$ to compare, and which cues are more relevant. 
We define this as a multi-task model with two sub-models with fully connected neural network layers $M_r$ and $M_y$. 
$M_y$ takes in the numeric cue differences $\hat{\bm{c}}^{y\gamma}$ and embedding representations (from the penultimate fully connected layer) of the emotion concepts $\hat{\bm{z}}_0^y$ and $\hat{\bm{z}}_0^\gamma$ to predict the emotion $\hat{y}$ heard in $x$. We determine which cues were more important by calculating an attribution explanation $\hat{\bm{w}}_c^{y\gamma}$ with layer-wise relevance propagation (LRP)~\cite{bach2015pixel}. These attributions are then concatenated on $\hat{\bm{c}}^{y\gamma}$ to determine the weighted cue differences $\hat{\bm{w}}_c^{y\gamma}$.
$M_r$ takes in $\hat{\bm{w}}_c^{y\gamma}$, $\hat{\bm{z}}_0^y$ and $\hat{\bm{z}}_0^\gamma$ to predict the cue difference relations $\hat{\bm{r}}_w^{y\gamma}$.
With the ground truth references, cue difference relations prediction can be trained using supervised learning.
Since the cue difference relations (lower, similar, higher) are ordinal, we employed the NNRank ordinal encoding~\cite{cheng2008neural} with 2 classes, such that lower = $(0,0)^T$, similar = $(1,0)^T$, higher = $(1,1)^T$, sigmoid activation, and binary cross-entropy loss for multi-label classification.
% ref: https://stats.stackexchange.com/a/324879, primary src: https://stackoverflow.com/a/38377488

\subsubsection{RexNet Model Summary}
RexNet consists of several modules to predict a concept and provide relatable explanations. 
Its primary task takes an input voice audio clip $\bm{x}$ to predict emotion concept $y$.
For explanations, by specifying contrast emotion concept $\gamma$ as input, the model generates explanations for the initial emotion concept $\hat{\bm{y}}_0$, contrastive saliency $\hat{\bm{\varsigma}}^{y\gamma}$, cue difference relations $\hat{\bm{r}}_w^{y\gamma}$, and cue difference importance $\hat{\bm{w}}_c^{y\gamma}$.
Each of these explanations and other absolute explanations can be provided to the end-user. 
% We describe an explanation user interface in the next section.

\subsection{Relatable Explanation User Interface}

\begin{figure}[t]
    \centering
    \hspace*{-0.5cm}
    \includegraphics[width=8.0cm]{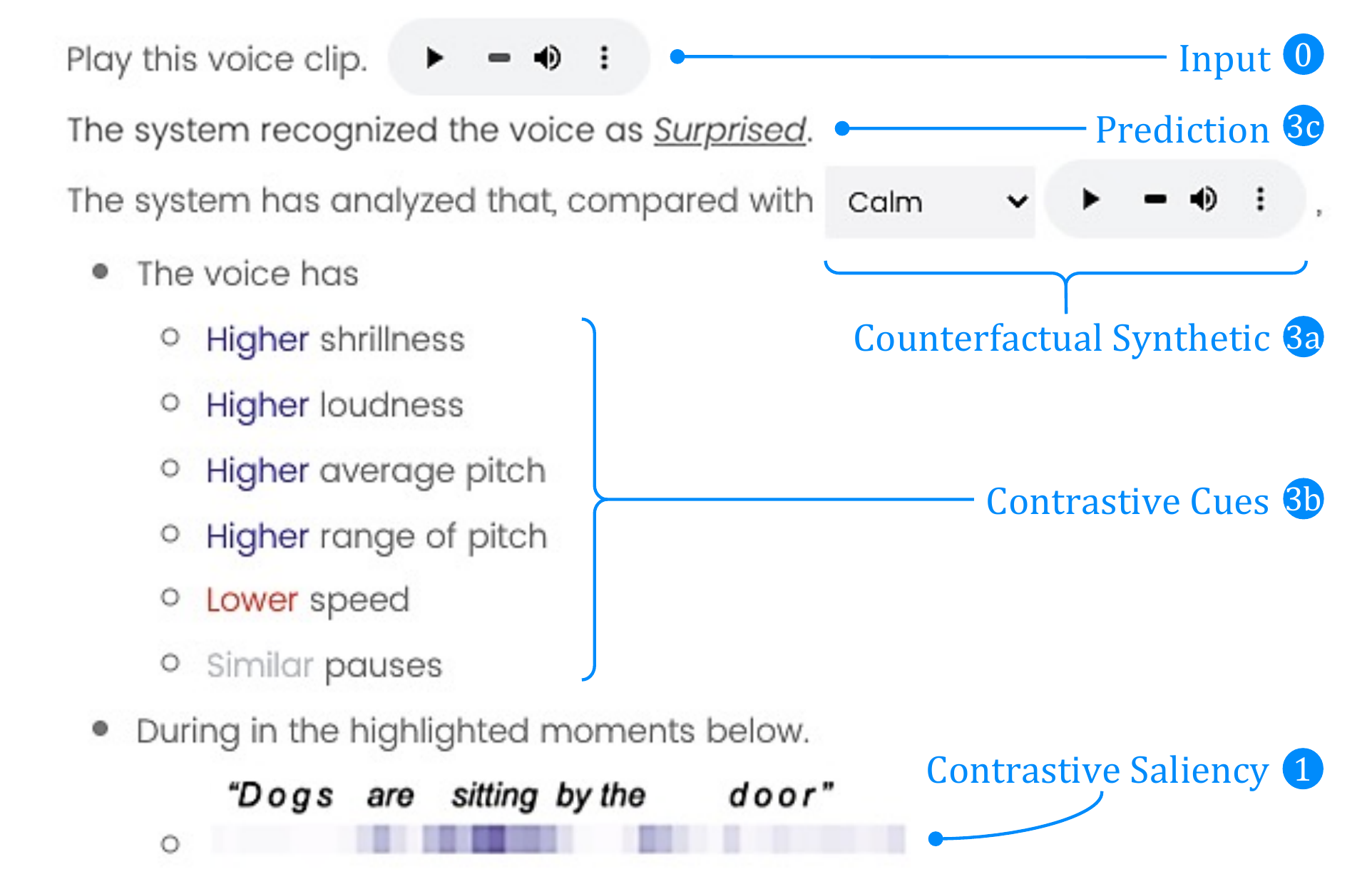}
    \vspace{-0.2cm}
    \caption{
    User interface of voice clip (RexNet step 0), predicted emotion (3c), and three relatable contrastive explanation types (1, 3a, 3b).
    The model prediction (3c) is omitted in the user study to test human-simulatability.
    }
    \label{fig:ui-screenshot}
    \Description{The user interface with five components from top to bottom: voice clip, system predicted emotion, Counterfactual Synthetic explanation, Contrastive Cues explanation, Contrastive Saliency explanation.}
    \vspace{-0.4cm}
\end{figure}

Fig. \ref{fig:ui-screenshot} shows the user interface with all relatable explanations.
After listening to a voice clip (Input), the user can read the model's recognition of the emotion (Prediction), listen to the voice as an alternative emotion (Counterfactual Synthetic), compare the cues between the target and counterfactual voice clips (Contrastive Cues), and see the salient moments in the heatmap (Contrastive Saliency).

\section{Evaluations}

We first evaluated the performance of our interpretable model, then conducted two user studies to evaluate the usage and usefulness of the contrastive explanations.
The first user study was formative to qualitatively understand usage, and the second was summative to measure the effectiveness of each explanation type.

\subsection{Modeling Study}

\subsubsection{Method}
We evaluated the model prediction performance and explanation correctness with several metrics (Table \ref{table:modeling_evaluation}).
We measured the accuracies of the initial and final predictions of emotion, and compared them against that of the baseline CNN model. 
% Performance was calculated with macro-average accuracy (of all classes).
Each explanation type was evaluated with different metrics due to their different forms.
We evaluated \textit{saliency maps} by the relevance of important features to the model prediction, and compared absolute and contrastive saliency. We employed the ablation approach of \cite{li2018tell} that identifies more important features as those that cause larger decreases in model performance when that feature is ablated.
We evaluated the faithfulness of \textit{counterfactual synthetics} with these metrics: 
1) reconstruction similarity $exp(-MSE(\bm{x}, \tilde{\bm{x}}^\gamma))$ between the input $\bm{x}$ and synthesized $\tilde{\bm{x}}^\gamma$, calculated with mean square error $MSE$, to determine how similar they are; 
% {\color{red}
% 2) the Levenshtein edit distance~\cite{levenshtein1966binary} between the sentences $STT(\bm{x})$ and $STT(\tilde{\bm{x}}^\gamma)$ recognized from a speech-to-text conversion\footnote{We used the Otter.ai speech-to-text API.} of the respective audio instances; 
% }
2) the identity classification accuracy to indicate whether the counterfactual voice sounds like the same actor portraying the original emotion; and
3) the emotion classification accuracy with respect to the contrast emotion.
We evaluated the correctness of \textit{cue difference relations} $\hat{\bm{r}_w^{y\gamma}}$ by comparing the inferred relations (i.e., higher, lower, similar) to the ground truth relations calculated from the dataset (e.g., see Table \ref{table:cueDifferences_vs_happy}). 
% We report the binary cross entropy error for each cue, i.e., $BCE(\hat{\bm{c}}^{y\gamma}, \bm{c}^{y\gamma})$, and the macro-average across all cues.
% We report the classification accuracy of the cues.
All multi-class metrics are reported with their macro-averages.

% Measures
% - performance of baseline emotion prediction: accuracy, confusion matrix (baseline CNN without XAI modules)
% - performance of initial emotion prediction: accuracy, confusion matrix
% - performance of final emotion prediction: accuracy, confusion matrix
% - relevance of saliency: GAIN ablation
% - relevance of contrastive saliency: GAIN ablation for y class (not macro-average)
% - faithfulness of counterfactual synthetic
%   - reconstruction error $MSE(\bm{x}, \tilde{\bm{x}}^\gamma)$
%   - edit distance between original sentence and sentence from speech2text recognition of synthesized speech audio, i.e., ... we used the Otter.ai API.
%   - classification accuracy $CCE(M(\bm{x}), \gamma)$ (categorical cross entropy)
% - correctness of cue differences: correctness with respect to the ground truth cue differences from the dataset

\begin{table}[t]
\small \setlength{\tabcolsep}{2.05pt}
\caption{
Evaluation results of model prediction performance and explanation correctness for RexNet with StarGAN or Counterfactual Samples and baseline models.
% RexNet models compared include the full model, and one trained with Counterfactual Samples (C.Samples) without StarGAN used in user studies. 
% We compared the full RexNet model and one trained with Counterfactual Samples (C.Samples) without StarGAN used in user studies.
Grey numbers calculated from definition. * same as Base CNN. 
}
\vspace{-0.2cm}
\begin{tabular}{llcccc}
\hline
                         &                                    & \multicolumn{4}{c}{Model}                                                                                               \\ \cline{3-6} 
Variable                 & Metric                             & {\color[HTML]{C0C0C0} Random} & \begin{tabular}[c]{@{}l@{}}Base\\ CNN\end{tabular} & RexNet & \begin{tabular}[c]{@{}c@{}}RexNet w/ \\ C.Samples\end{tabular} \\ \hline
\begin{tabular}[t]{@{}l@{}}Initial\\ Concept $\hat{y}_0$\end{tabular}     & \begin{tabular}[t]{@{}l@{}}Emotion accuracy\\ (8 classes)\end{tabular}       & {\color[HTML]{C0C0C0} 12.5\%} & 75.7\%   & 79.5\% & 78.8\%                                                              \\
\begin{tabular}[t]{@{}l@{}}Final\\ Concept $\hat{y}$\end{tabular}           &\begin{tabular}[t]{@{}l@{}}Emotion accuracy\\ (8 classes)\end{tabular}       & {\color[HTML]{C0C0C0} }       &          & 78.5\% & 77.4\%                                                              \\ \arrayrulecolor{lightgray}\hline
\begin{tabular}[t]{@{}l@{}}Absolute\\ Saliency $\hat{\bm{s}}^y$ \end{tabular}        & \begin{tabular}[t]{@{}l@{}}Ablated accuracy\\ decrease\end{tabular}          & {\color[HTML]{C0C0C0} }       &          & 14.9\% & 16.7\%                                                              \\
\begin{tabular}[t]{@{}l@{}}Contrastive\\ Saliency $\hat{\bm{\varsigma}}^{y\gamma}$\end{tabular}    & \begin{tabular}[t]{@{}l@{}}Ablated accuracy\\ decrease\end{tabular}          & {\color[HTML]{C0C0C0} }       &          & 13.7\% & 16.3\%                                                              \\ \arrayrulecolor{lightgray}\hline
\begin{tabular}[t]{@{}l@{}}Counterfactual\\ Synthetic $\hat{\bm{x}}^\gamma$\end{tabular} & \begin{tabular}[t]{@{}l@{}}Reconstruction\\ similarity\end{tabular}          & {\color[HTML]{C0C0C0} }       &          & 0.553  & {\color[HTML]{C0C0C0}1}                                        \\
                        %  & {\color{red}Words Levenshtein distance}         & {\color[HTML]{C0C0C0} }       &          & ?      & {\color[HTML]{C0C0C0} 100\%}                                        \\
                         & \begin{tabular}[t]{@{}l@{}}Identity accuracy\\ (24 classes)\end{tabular}     & {\color[HTML]{C0C0C0} 4.2\%}  &          & 60.2\% & 96.2\%                                                              \\
                         & \begin{tabular}[t]{@{}l@{}}Emotion accuracy\\ (8 classes)\end{tabular}       & {\color[HTML]{C0C0C0} 12.5\%} &          & 30.7\% & {\color[HTML]{C0C0C0}75.7\%*}                                                             \\ \arrayrulecolor{lightgray}\hline
\begin{tabular}[t]{@{}l@{}}Cue Difference\\ Relation $\hat{\bm{r}}^{y\gamma}_w$\end{tabular}   & \begin{tabular}[t]{@{}l@{}}Cue accuracy\\ (3 classes, 6 labels)\end{tabular} & {\color[HTML]{C0C0C0} }       &          & 71.9\% & 71.6\%                                                              \\ \arrayrulecolor{black}\hline
\end{tabular}
\label{table:modeling_evaluation}
\vspace{-0.3cm}
\end{table}

\subsubsection{Results}

We split the dataset into 80\% training and 20\% test. Table \ref{table:modeling_evaluation} reports the test results.
Training with the explainable modules helped RexNet to achieve higher emotion accuracy than the base CNN (79.5\% vs. 75.7\%). Though the final emotion accuracy is slightly lower than the initial emotion prediction (78.5\% vs. 79.5\%), this is expected since interpretability typically trades-off accuracy~\cite{gunning2019darpa}.
The ablated accuracy decrease indicates that the saliency pixels are somewhat important. Contrastive Saliency has slightly less importance than Absolute Saliency (13.7\% vs. 14.9\%), because the former excludes pixels that are commonly important for all classes.
Counterfactual synthesis was moderately successful, achieving reasonable reconstruction similarity (reconstruction error MSE = 0.680), good speaker re-identification (60.2\% compared to 4.2\% random chance), and somewhat recognizable emotion which is significantly better than random chance (30.7\% vs. 12.5\%).
The predictions of cue difference relations were good (71.9\%).

Although the counterfactual synthesis accuracy was better than chance, it is still too low to be used by people. Hence, we evaluated instead using Counterfactual Samples (C.Samples), which uses actual voice clips %from RAVDESS~\cite{livingstone2018ryerson} 
corresponding to the same voice actor (identity), same speech words, but different portrayed contrast emotion. As expected, the identity and emotion accuracies are higher for Samples than Synthetics, but the other performances were comparable.

In the next step, we investigate the usage and usefulness of each explanation type.
The focus is on the interactions and interface, rather than investigating whether each explanation as implemented is good enough.
Therefore, we select instances with correct predictions and coherent explanations for the user studies.
Since the Counterfactual Synthesis performance is limited, we use Counterfactual Samples to represent counterfactual examples instead.

\subsection{Think-Aloud User Study}
We conducted a formative study with the think-aloud protocol to understand how people
1) naturally infer emotions without AI assistance, and
2) use or misunderstand various explanations.

\subsubsection{Experiment Method and Procedure}
We recruited 14 participants from a university mailing list. They were 3 males, 11 females, with ages between 21-40 years old.
We conducted the study via an online Zoom audio call. 
The experiment took 40-50 minutes and each participant was compensated with a \$7.43 USD coffee gift card.
% \$10 SGD
The user task is a human-AI collaborative task for vocal emotion recognition. Given a voice clip, the participant infers the portrayed emotion with or without AI prediction and explanation. We provided 16 voice clips of 2 neutral sentences\footnote{Neutral sentences: "dogs are sitting by the door" and "kids are talking by the door"} intoned to portray 8 emotions.
We selected only correct system predictions and explanations, since we were not investigating the impact of erroneous predictions or misleading explanations.
The study contains 4 explanation interface conditions: Contrastive Saliency only, Counterfactual Sample voice examples only, Counterfactual Sample and Contrastive Cues, and all 3 explanations together (Fig. \ref{fig:ui-screenshot}).

The procedure is: read an introduction, consent to the study, complete a guided tutorial of all explanations (regardless of condition), and start the main study with multiple trials of a vocal emotion recognition task.
To limit the participation duration, each participant completes three trials, each trial randomly assigned to an explanation interface condition.
For each trial, the participant listened to a voice clip, and gave an initial label of the emotion. 
On the next page, the participant was shown the system’s prediction with (or without) explanation based on the assigned condition.
She could then revise her emotion label if she changed her mind.
We used the think-aloud protocol to ask participants to articulate their thoughts as they examined the audio clip, prediction and explanations. We also asked them about their perceptions using the interface, and any suggestions for improvement.
We describe our findings next.

\subsubsection{Findings}
We performed thematic analysis on the recorded audio to determine key themes (bolded).
We describe our findings in terms of our research questions of how users innately infer vocal emotions, and how they used each explanation type.
When inferring on their own (without XAI), participants would 
\textbf{focus on specific cues} to
\textit{“check the intonations [pitch variation] for decision”} [Participant P12], infer a Sad emotion based on the \textit{“flatness of the voice”} [P04], or \textit{"use shrillness to distinguish between fearful and surprise"} [P01].
Participants also relied on \textbf{changes in tone}, which we had not modeled. For example, a rising tone \textit{“sounds like the man is asking a question”} [P02], \textit{“the last word has a questioning tone”} [P03] helped participants to infer Surprise. The latter case identified the \textbf{most relevant segment}.
In contrast, a \textit{"tone going down at the end of sentence"} helped P01 infer Sad.
Some participants \textbf{mentally generated} their own examples to \textit{"imagine what neutral sound like and compare against it"} [P05].
These unprompted behaviors suggest the relevance of saliency, counterfactual, and cue explanations.

The usage of explanations was mixed with some benefits and some issues.
In general, participants could understand the {\color[HTML]{008BFB}Saliency} maps. P09 saw that \textit{“the highlight parts are consistent with my judgment for important words”}, referring to ‘talking’ being highlighted.
However, several participants had issues with saliency maps. 
There were some cases with highlights that spanned across multiple words and included highlighting spaces. 
P08 felt that saliency \textit{“should highlight all words”}, and P14 \textit{“would prefer the color highlighted on text”}. This lack of focus made P13 feel that \textit{“the color bar is not necessary”}.
Regularizing the explanation to prioritize highlighting words and penalize highlighting spaces can help align the explanations with user expectations and improve trust \cite{ross2017right}
Next, P11 thought that \textit{“the color bar reflects the fluctuation of tone”}. While plausible, this indicates the risk of misinterpreting technical visualizations for explanations.
Finally, P12 \textit{“used the saliency bar by listening to the highlighted part of the words, and try to infer based on intonation. But I think the highlighting in this example is not accurate”}. This demonstrates \textbf{causal oversimplification} by reasoning with one factor rather than multiple factors~\cite{damer2012attacking,lim2011design}.

Many participants found \textbf{{\color[HTML]{008BFB}Counterfactual} samples “intuitive”}.
P11 could \textit{“check whether it’s consistent with my intuition”} by mentally comparing the similarity of the target audio clip (sad) with clips for other suspected emotions (neutral, sad, happy). Unfortunately, her intuition was somewhat flawed, since she inferred Neutral which was wrong.
P12 found counterfactuals \textit{“helpful to have a reference state, then I will also check the intonations for my decision.”}
Conversely, some participants felt counterfactual samples were not helpful.
P06 felt that the \textit{“clips [neutral and calm] are too similar”}. Had she received deeper explanations with saliency map or cue differences, she would have had more information about where and what the differences were, respectively.

{\color[HTML]{008BFB}Cues} were used to check semantic consistency.
P04 used cues to \textit{“confirm my judgment”} and found that the \textit{“low shrillness [of Sad] is consistent with my understanding.”}
% P06 reported that \textit{“cues mainly help me confirm my choice.”}
However, some participants perceived inconsistencies.
P13 thought that \textit{“some cue descriptions were not consistent with my perception”}, and disagreed with the system that Speaking Rate was similar for the Happy and Surprised audio clips.
Along with the earlier case of P06, this suggests differences in \textbf{perceptual acuity} of cues between the user and system.% to distinguish cue similarity.

Finally, some participants felt that {\color[HTML]{008BFB}Counterfactual samples} were more useful than {\color[HTML]{008BFB}Contrastive Cues}.
P11 found that \textit{“the comparison voice part is more helpful than the text part, though the text part is also helpful to reinforce my decision.”}
This could be due to cognitive load and differences between mental \textbf{dual processing}~\cite{kahneman2011thinking}. 
Many participants considered the audio samples \textit{“quite intuitive”} [P04]. They used System 1 thinking which is fast, though they did not articulate why this was simple.
In contrast, they found that \textit{“it’s hard to describe or understand the voice cue patterns”} [P04]. P10 felt that \textit{“compared with [audio] clips, cue pattern is too abstract to use for comparison.”}
This requires slower System 2 thinking.
Another possible reason is that the audio clip has higher information bandwidth than the 6 verbally presented semantic cues. Participants can perceive the \textbf{gestalt}~\cite{koffka2013principles} of the audio to make their inferences.

\subsection{Controlled User Study}
Having identified various benefits and usages of contrastive explanation, we next conducted a summative controlled study to understand:
1) how well participants could infer vocal emotions on their own, and with model predictions and explanations, and
2) how various explanations affect perceived helpfulness.

\subsubsection{Experiment Design and Apparatus}
We conducted a between-subjects experiment with XAI Type as the independent variable with 5 levels of explanations (None, Contrastive Saliency, Counterfactual Sample, Counterfactual + Contrastive Cues, and Saliency + Counterfactual + Cues). 
The user task is to label the portrayed emotion in a voice clip with feedback from the AI in one of the XAI Types. We included emotion as a random variable with 8 levels 
% (Neutral, Calm, Happy, Fearful, Surprise, Sad, Disgust, and Angry)
. Having many emotions helps to make the task more challenging to test.
Fig. \ref{fig:ui-screenshot} shows the UI with all explanations together, and others are shown in Supplementary Figs. \ref{appendix-figure:trial_none}-\ref{appendix-figure:trial_all}.
For dependent variables, we measured decision quality (emotion label correctness and confidence), understanding of cue differences, task times, decision confidence, and perceived system helpfulness.
Labeling correctness was measured with a “balls and bins” question~\cite{goldstein2014lay} that elicits the probability of multiple labels.
Cue difference understanding was measured per cue with a multiple choice question for the cue difference relation between a randomly selected contrast emotion label and the target voice clip.
Task times were logged for different pages.
Perceptions were measured as ratings on a 7-point Likert scale ($-3 =$ Strongly Disagree, $+3 = $ Strongly Agree).
% {\color{blue}We asked two open-ended text questions to justify perceived helpfulness or unhelpfulness, and to understand how the explanation was used. This was posed only twice to limit fatigue.}
We asked two text questions about the rationale for perceived helpfulness and how the explanation was used. This was posed only twice to limit fatigue.
See Supplementary Figs. \ref{appendix-figure:trial_none}-\ref{appendix-figure:trial_postxai} for the survey.

\subsubsection{Experiment Procedure}
The participant reads an introduction, consents to the study, reads a short tutorial about the explanation interfaces, and completes a screening test of audio equipment, auditory acuity, and UI understanding (Supplementary Figs. \ref{appendix-figure:slider}-\ref{appendix-figure:tutorial_comparisonSample}), where she: 
a) listens to a voice clip and chooses the correct spoken words, 
% We cannot enforce which listening devices participants used, but can ensure that they can hear the sounds.
% The auditory requirement does not need to be very strict, since perceiving vocal emotions is an everyday task for able-bodied people under various environmental conditions.
b) reads a saliency map and identifies important words, and 
c) identifies easy cue differences between two voice clips. 
% The screening tests the participant's audio equipment and auditory acuity.

After passing screening (with all questions correct), the participant is randomly assigned to an XAI Type and commences a practice session.
Similar to \cite{lim2009and}, we conducted the practice session to enable the participant to learn from any model explanations how the system predicts the emotion. 
% She sees both the explanation and prediction together to learn their relationship. 
She is encouraged to study these cases carefully, since she will not see the correct predictions later in the main study.
The practice session comprises 8 trials, where each trial has three pages:
i) \textit{Pre-AI} to listen to a voice clip, label the emotion without AI assistance.
% and answer questions about cue differences.
We assess the labeling correctness here to estimate the Participant Unaided Skill, i.e., whether the participant has above- or below-average skill in recognizing vocal emotions.
ii) \textit{Post-XAI} to read any explanation feedback (without seeing the system prediction, label the emotion (again), and answer questions about cue difference understanding, and perceived ratings.
iii) \textit{Review} to examine the correct emotion label (same as the system prediction) with any AI explanations, and the participant's previous answer, and write any free-form notes (open text).

After the practice session, the participant engages in the main study with the same XAI Type in two sessions with 8 trials each and a break in-between.
Each trial is presented on one page where the participant:
i) listens to the voice clip,
ii) views any explanation feedback,
iii) labels the emotion, and
iv) rates perceptions.
This evaluates human-simulatability~\cite{lim2009and,lipton2018mythos} by deeply testing the participant's understanding
% \footnote{While this does not represent real-world usage of XAI to verify system predictions, it avoids the confound of participants copying a displayed prediction.} 
to apply explanations to new instances.
To control any fatigue effects due to the moderate number of trials, we randomized the order of instances.
We asked the rationale questions randomly in one trial per main session.
The participant is incentivized to be fast and correct with a maximum \$0.50 USD bonus for completing all trials within 8 minutes. The bonus is pro-rated by the number of correct emotion labels. Maximum bonus is \$1.00 for two sessions over a base compensation of \$3.00 USD. 
The participant ends with answering demographic questions.
\begin{figure}[t]
    \centering
    % \hspace*{-0.5cm}
    \includegraphics[width=7.4cm]{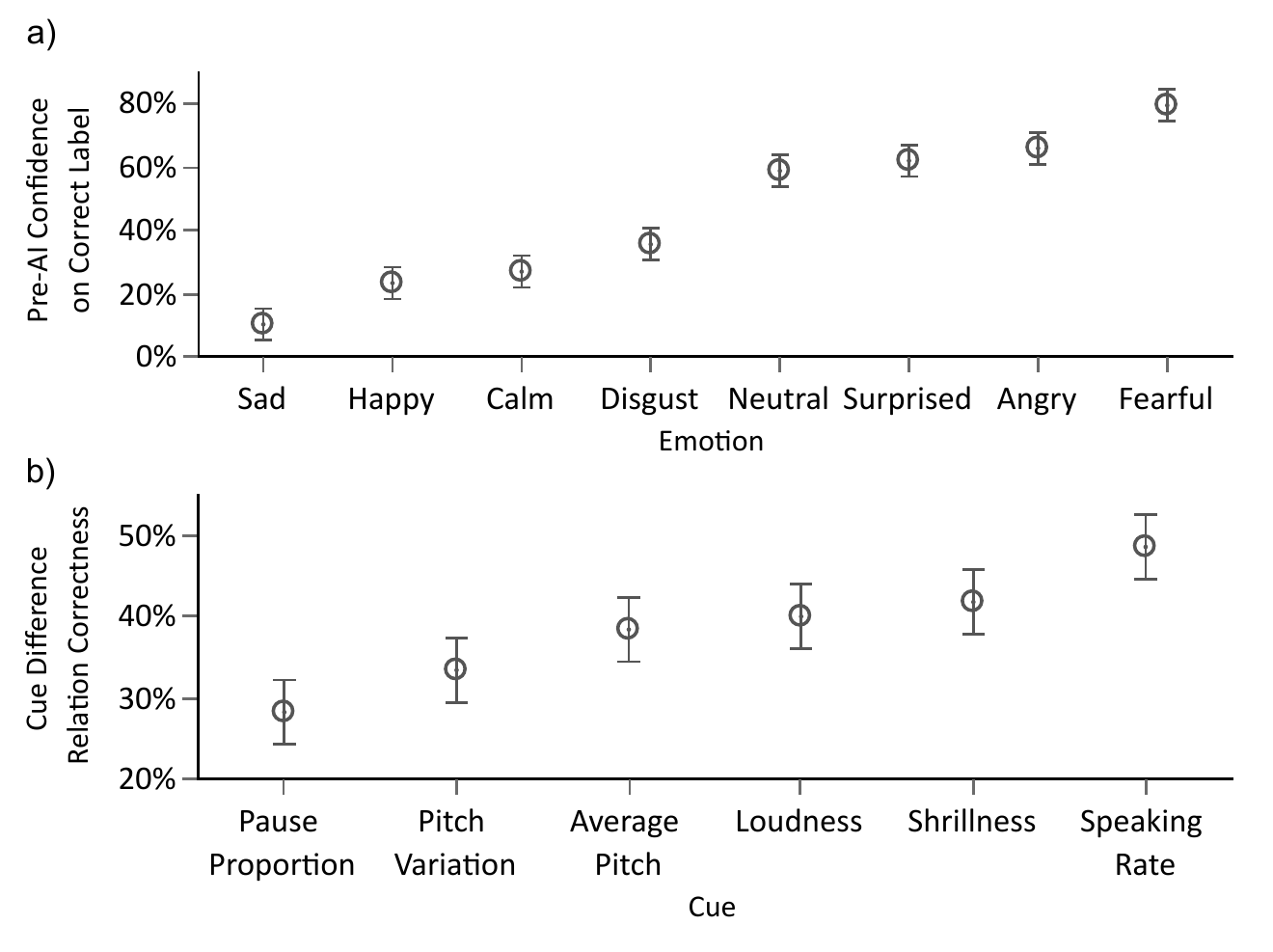}
    \vspace{-0.4cm}
    \caption{
    Participant audio perception skill across emotions and vocal cues.
    a) User confidence on the correct label to recognize emotions on their own without AI. 
    b) User correctness of perceiving cue differences between two voice clips.
    }
    \label{fig:quantitative-summary-results}
    \Description{a) Barchart of pre-XAI labeling confidence on correct label for 8 emotions. Sad has the lowest mean value while Fearful has the highest. b) Barchart of cue difference relation correctness for 6 cues. Pause proportion has the lowest mean value while Speaking Rate has the highest.}
    \vspace{-0.45cm}
\end{figure}

\subsubsection{Statistical Analysis and Quantitative Results}

We recruited 175 participants from Amazon Mechanical Turk with high qualifications ($\geq5000$ completed HITs with \textgreater 97\% approval rate). They were 52.0\% male, with ages 21-70 (Median = 36).
Participants took 27.4 minutes (median) to complete the survey. 
We excluded 14 participants who completed the survey without playing any voice clips. 

% \begin{figure*}[h!]
%     \centering
%     % \hspace*{-0.5cm}
%     \includegraphics[width=15.8cm]{figures/quantitative-summary-results.pdf}
%     \caption{
%     Participant audio perception skill across emotions and vocal cues.
%     a) User confidence on the correct label to recognize emotions on their own without AI. 
%     b) User correctness of perceiving cue differences between two voice clips.
%     }
%     \label{fig:quantitative-summary-results}
%     % \vspace{-0.2cm}
% \end{figure*}

\begin{figure*}[t]
    \centering
    % \hspace*{-0.5cm}
    \includegraphics[width=15.8cm]{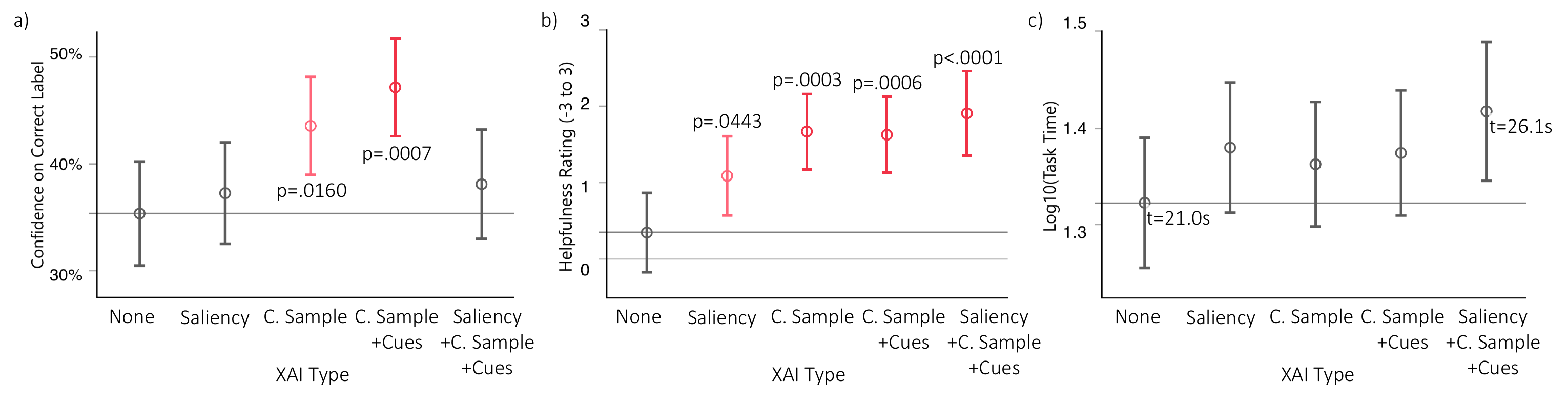}
    \vspace{-0.3cm}
    \caption{
    Results of relatable explanations on a) labeling correctness, b) perceived helpfulness, and c) task time for AI-assisted emotion recognition.
    Significant difference from None are indicated with p-values.
    XAI Types have various explanation combinations: None (no explanations), Contrastive Saliency, Counterfactual Sample (C.Sample), and Contrastive Cues.
    }
    \label{fig:quantitative-results}
    \Description{a) Barchart of post-XAI labeling confidence on correct label for 5 XAI conditions. C.Sample+Cues is significantly higher than None and C.Sample is marginally higher than None. b) Barchart of Helpfulness Rating for 5 XAI conditions. C.Sample, C.Sample+Cues and Saliency+C.Sample+Cues is is significantly higher than None while Saliency is marginally higher than None. c) Barchart of Response Time for 5 XAI conditions. No significance found. }
    \vspace{-0.3cm}
\end{figure*}

For each dependent variable, we fit a linear mixed-effects model with XAI Type, Emotion, Voice Clip, Participant Unaided Skill and Trial Number as main fixed effects, and Participant as random effect. We did not find any significant interaction effects. See Supplementary Table \ref{table:statModelDetails} for details. 
We report significant results at a stricter significance level (p<.005) to account for multiple comparisons.
% We supplement some of our quantitative analyses with relevant quotes from participant rationales where relevant.
% We focus on results regarding XAI Type.

Regarding emotion labeling in the Pre-AI Practice Trials, participants recognized some emotions better than others (Fig. \ref{fig:quantitative-summary-results}a) and perceived different cues with varying accuracies (Fig. \ref{fig:quantitative-summary-results}b), indicating that speaking rate and shrillness could be most verifiable in explanations, while pause proportion and pitch variation may be least. 
Furthermore, there was a wide range of average correctness among participants (M=49.9\%, SD=18.3\%), so we divided participants by whether they had above- or below-average unaided skill.

Analyzing the Main Trials, we found varying performances and perceptions due to different XAI Types (Fig. \ref{fig:quantitative-results}).
Although participants may select a wrong emotion label as most likely, they may still select the correct label with low confidence in the balls and bins question. Hence, we analyzed the Confidence on Correct Label to determine the participant's decision quality. Results were similar when analyzing with labeling correctness. Fig. \ref{fig:quantitative-results} shows that providing Counterfactual Sample voices with Cues (C.Sample + Cues) were most effective and significantly better than not providing any explanation (None), p=.0007. Omitting the cues (C.Sample) led to a decrease in decision quality such that the difference from None was marginal, p=.0160. Providing Contrastive Saliency explanations did not help to improve decision quality, and, surprisingly, neither did providing all explanations combined together. % (Saliency + C.Sample + Cues).
All XAI Types were rated as more helpful than None, though Saliency was only marginally so (p=.0443).
All participants were equally confident (p=n.s.) in their emotion labels across XAI Types (Median=2 on -3 to 3 Likert scale).
There was no difference in task time to label the emotions, though the most complex explanations (Saliency + C.Sample + Cues) was only 4.1 sec longer than None (26.1 vs. 21.0s).

\subsubsection{Qualitative Results}
We report why participants found specific XAI Types helpful or unhelpful and how they used them.
Some participants depended on their own ability than rely on explanations, e.g.,
\textit{"I don't think that the [Saliency] explanation information is helpful. I think that the voice is all you really need to be able to determine an emotion."} [P169], \textit{"I didn't [use C.Sample] for the most part. I trust my own instincts."} [P54], \textit{"I think [Saliency + C.Sample + Cues] took longer than just listening to the clip ... I glance over it, but it doesn't affect my decision as much"} [P121].

Some participants struggled to use the {\color[HTML]{008BFB}Contrastive Saliency} map. P26 found it \textit{"difficult to parse ... hard to analyze it by eye)"}. Errors in the Saliency explanations also led to distrust, as described by P8 that \textit{"the highlighted moments for Fearful don't match well with [the] voice".} Conversely, the sophistication of the explanation led to over-trusting, with P146 mentioning that it was \textit{"helpful to view the color bar to determine which part has the most importance"}, yet, this shallow interpretation led to him labeling wrongly.
P117 commented that she was \textit{"unable to listen to different ratings the system has given to each emotion"}, indicating her desire to hear other samples.

{\color[HTML]{008BFB}Counterfactual Sample} explanations were more appreciated and marginally effective in improving decision quality.
P38 felt that the \textit{"emotion in the clip is very clearly anger and it helped to hear the system show me what this voice would sound like when angry"}; thus, she was matching samples by their perceived similarity. Similarly, P132 \textit{"first made my own judgment to narrow down the possible emotions, then listen to those emotions. I rate the one that matches the highest."} In contrast, P14 felt that C.Sample was \textit{"helpful to tell the difference between the neutral and calm voice"} and \textit{"tried to see if there was a change in inflection or speed"}.
P103 felt that \textit{"it is slightly far away from the sample clip, every single one of them"}, suggesting that he would appreciate Counterfactual Synthetics which would be generated to be more similar.
Finally, P54 demurred that \textit{"the explanation information doesn't elaborate at all why it's giving that determination, so it's mostly not helpful"}; this indicates the need for deeper semantic explanations which C.Sample + Cues provides.

Instead of manually perceiving similarities or differences in voice clips, participants could read the cue differences in the {\color[HTML]{008BFB}Counterfactual Sample + Cues} explanation.
Their analytical understanding improved, as demonstrated in the vocabulary of their rationalization; e.g., P119 had a \textit{"better sense of the speaker's pitch, loudness"}.
% , P90 \textit{"utilized the explanation information to eliminate my suspicions or questions about a certain speed, pitch, loudness or shrillness of an emotion"}.
The semantic knowledge provided by cues also helped to reduce cognitive burden, e.g., P168 \textit{"used the information to confirm something I feel ambiguous about or just to make a guess and not have to spend so much effort deciding between guesses."} 
Specifically, cues helped to focus participants' analyses, e.g., P90 found the explanation \textit{“helpful in letting you figure out what qualities to try to isolate in the voice clip to decide on where it learns in terms of emotion.”}

Finally, although participants perceived {\color[HTML]{008BFB}Saliency + Counterfactual Sample + Cues} as helpful, it did not improve decision quality.
Participants rationalized the explanations by describing its various components separately, e.g.,
\textit{"it helps pinpoint what parts to listen to"} [P31, Saliency], 
\textit{"the sample clips for each emotion are [helpful]"} [P163, C.Sample], 
% \textit{"helpful because of the high pitch of the tone"} [P108, Cues], 
\textit{"compare the voices and the levels (like shrillness and pitch)"} [P15, Cues].
However, no one explicitly described multiple components together, and there were few explicit descriptions about the saliency map.
Perhaps, participants could not focus on specific explanation details. P167 was \textit{"not sure how to apply cross the broad"}, suggesting an issue with information overload.

\subsection{Summary of Results}
We summarize the results from our three evaluation studies.
The modeling study showed that RexNet provides relevant Saliency explanations, accurate Contrastive Cues explanations, and promising Counterfactual Synthetics. These explanations helped to improve RexNet's performance over the base CNN.
The think-aloud user study showed that RexNet explanations align with how users innately perceive and infer vocal emotions, validating the XAI Perceptual Processing framework. We identified limitations in user perception and reasoning that led to some interpretation issues.
The controlled user study showed that some relatable explanations can improve decision quality without sacrificing task time, especially Counterfactual Samples with semantic Cues. Saliency visualization is too technically sophisticated to be useful, and combining it with Counterfactual Samples and Cues could improve the perceived helpfulness, but also confuse or distract participants to decide poorly.

\section{Discussion}
Having evaluated our framework for relatable explainable AI, we discuss their usefulness, improvements to our approach and experiment, implications for human-centric XAI, and generalization.

\vspace{-0.1cm}
\subsection{Usefulness of Relatable Explanations}
Our proposed XAI Perceptual Processing Framework and RexNet architecture unifies different explanations towards relatability. We have rationalized their relevance based on cognitive theories, demonstrated their benefit to improving model prediction performance, and partially validated their usefulness in user studies. 
We discuss takeaways for XAI developers to design relatable explanations.

The effectiveness of Counterfactual + Cues explanations indicates the value of augmenting example-based explanations with semantic information.
However, we found that saliency explanations had limited usefulness. 
Furthermore, adding Saliency to Counterfactual + Cues nullifies any benefits of the latter.
Our findings contradict those by Wang et al. \cite{wang2021explanations} that attribution explanations were more useful than counterfactuals, possibly due to the difference of interpreting structured or unstructured data.
Despite many XAI techniques being developed as saliency maps (e.g., \cite{Zhang2018VisualIF}), there have been recent calls to develop more meaningful explanations of image prediction tasks~\cite{ghorbani2019towards}. 
Thus, saliency maps should not be used or need to be made more precisely correct (e.g., through model training or regularizations) and more semantically meaningful.

To address the weaknesses of some relatable explanations, we discuss ways to further improve their effectiveness.
Using counterfactual synthetics, instead of counterfactual samples would refine the difference between the example and target, so this may focus the user's attention to more meaningful differences and improve discriminating between concepts.
Moreover, our current approach identifies one set of cue differences across multiple salient locations. Instead, different cue sets can be associated with specific highlights in the saliency map. This can provide more semantics to various parts of a saliency map, to indicate why particular regions were important, and improve the usefulness of saliency maps.

\vspace{-0.1cm}
\subsection{User Evaluation of Relatable Explanations}
We chose to evaluate with vocal emotion recognition since it is an everyday task that is feasible to test with lay users. However, most people are already innately skilled in this, so this diminishes their need for AI or XAI to help them. 
Conversely, relatable explanations may be more useful for more analytical tasks and applications with more explicit domain knowledge (e.g., engine noise diagnosis).

% We had evaluated explainable AI (XAI) towards improving human-AI decision quality, while XAI is also commonly used to improve human trust in AI decisions. In real-world deployments, the AI should be performing most tasks automatically; for example, the emotion should be automatically predicted by the model, instead of only supplying explanations and having the user do the recognition.
% However, conducting experiments with the latter, more realistic, use case can suffer from experimental confounds. Showing the model prediction with the explanation and asking participants to verify the label, is prone to participants just copying the predictions. In a pilot study, we found that a system with prediction-only was more useful than explainable systems to improve labeling. This has confounded user understanding, trust, and convenience. On the other hand, our experiment design carefully evaluated understanding as human-simulatability, though at the cost of ecological validity.

We had identified several potential confounds --- fatigue, skill at recognizing emotions, participants copying system predictions, learning effects from exposure to prior XAI versions --- and discuss how we mitigated them.
1) We controlled for fatigue by: 
a) providing breaks between sessions,
b) randomizing instances across trial numbers.
We checked for fatigue by measuring:
a) repeatedly identical responses (no participants were disqualified),
b) decreases in labeling correctness over trials (no significant difference).
2) We controlled for recognition skills by measuring labeling performance without XAI (Pre-AI) and analyzed our results with that as a factor.
3) 
A more realistic use of AI is for it to make predictions and the user would verify its decision. However, in a pilot study evaluating with this task, we found that participants may copy the prediction rather than study the explanation, thus leading to over-trusting~\cite{yin2019understanding} and diminishing the usefulness of explanations to improve decision quality.
We mitigated copying by evaluating with a human-simulatability task, instead of a predicted label verification task, though this trades-off some ecological validity.
4) We mitigated learning effects by designing the experiment as between-subjects, otherwise, participants may exploit new knowledge in subsequent experiment conditions (with weaker explanations).

% \subsection{Extension of relatable explanations for vocal emotion recognition}
\vspace{-0.1cm}
\subsection{More relatable vocal emotion explanations}
This work is the first to explore relatable explanations for vocal emotion prediction, with an initial set of cues and adequate explanation accuracy.
% In this work, we have focused on recognizing emotions by their verbal expressions~\cite{juslin2001impact}. 
Future work can leverage other vocal stimulus types and prosodic attributes~\cite{lausen2020emotion}, such as non-verbal expressions, affect bursts, and lexical information. In particular, we learned that participants focus on the change in tone in voices to infer emotion, so this should be included as a vocal cue.
Counterfactual synthesis accuracy can be improved by using newer generators, such as Sequence-to-Sequence Voice Conversion \cite{kameoka2020convs2s}, StarGAN-VC v2~\cite{kaneko2019stargan}.
Though generated from a unified architecture, the explanations still had some inconsistencies. Annotating and debiasing explanations ~\cite{li2018tell,zhang2020debiased} could help to align explanations with user expectations~\cite{ross2017right} and improve the coherence between explanation types.
Contrastive Cue relations were encoded as a table, but they could be represented as another data structure (e.g., decision trees or causal graphs) to better fit human mental models.
Finally, further testing could evaluate the usage and usefulness of predictions and explanations in in-the-wild applications~\cite{lim2013evaluating}, such as with smart speakers (e.g., Amazon Echo)~\cite{maharjan2019hear}, smartphone digital assistants for mental health or emotion monitoring~\cite{wang2014studentlife, ben2015next}, or AI coaching for call center employees~\cite{gorrostieta2018attention,luo2021artificial,petrushin1999emotion}.

% \subsection{Improvement of semantic relatability for human-centric explainable AI}
\vspace{-0.1cm}
\subsection{Relatability for human-centric XAI}
Although many XAI techniques have been recently developed, many remain too technical, or focus on supporting data scientists and machine learning model developers.
Instead, there is a growing call to support different stakeholders and less technical users~\cite{ehsan2018rationalization, liao2020questioning, cheng2019explaining}
Towards this end, we have studied human perception and cognition to determine new requirements for XAI.
Miller had argued for contrastive and counterfactual explanations based on philosophical and psychological principles~\cite{miller2019explanation}. This was extended by Wang et al. to identify human reasoning pathways that can be supported by specific XAI techniques~\cite{wang2021explanations}.
We extend these perspectives by identifying a broader requirement that explanations need to be relatable and contextualized to be more meaningfully interpreted. 
Specifically, we supported four criteria for relatability: contrastive concepts, saliency, counterfactuals, and associated cues.
Extending our work, explanations can be made more relatable by providing for other criteria such as: social proof~\cite{cialdini1999compliance,miller2019explanation}, narrative stories~\cite{segel2010narrative} or rationalizations~\cite{ehsan2018rationalization}, analogies~\cite{gentner2012analogical}, user-defined concepts~\cite{kim2018interpretability, zhou2018interpretable, ghorbani2019towards}, and plausible explanations~\cite{ross2017right}.
Moreover, human cognition has natural flaws, like cognitive biases and limited working memory. 
XAI should include designs and capabilities to mitigate cognitive biases~\cite{wang2019designing}, moderate cognitive load~\cite{abdul2020cogam}, and accommodate information handling preferences~\cite{wang2021show}.
Relatable explanations may need to account for these human factors to communicate why they may deviate from human reasoning.
% Users may prefer systems to be less perfect to be more relatable (e.g., not updating too fast~\cite{yablonski2020laws})
% - Need to consider other cognitive effects to make it more human-like: perceiving sequential data (e.g., audio) is biased by timing effect (e.g., primacy, and recency); some characteristics are more noticeable by people too (e.g., volume vs. pitch [cite]). ML can be regularized to align more with human perception.

The XAI Perceptual Processing Framework was inspired by human perceptual reasoning, rather than higher-level cognition. The latter is relevant for complex decision-making tasks, such as doctors' reasoning with disease models, which are specific cases of causal structural models. Wang et al. proposed the XAI Reasoning Framework based on human reasoning processes~\cite{wang2019designing}, but this was not explicitly implemented in a single machine learning architecture.
The Intelligibility Toolkit~\cite{lim2010toolkit} provided an API to automatically generate explanations to a taxonomy of questions~\cite{lim2009assessing,liao2020questioning}, but this was not implemented for deep learning.
Future work can explore a meta-model that combines perceptual and reasoning faculties for more complex, human-like model explanations.

\vspace{-0.1cm}
\subsection{Generalization of Relatable XAI}
Although we implemented RexNet for the application of vocal emotion recognition, the XAI Perceptual Processing Framework is generalizable to other audio and visual prediction applications.
Other audio applications include equipment monitoring via vibrations~\cite{Wegerich2003NonparametricMO}, and heart murmur diagnosis~\cite{Reed2004HeartSA}.
1) Saliency can be highlighted on a spectrogram of vibration signals for trained engineers to interpret, or highlighted on auscultation diagrams for clinicians.
2) Counterfactual samples can be archetypal sounds of engine failure, specific heart disease (e.g., crescendo-decrescendo murmur in aortic stenosis), etc.
3) Cues can be the sound profiles, such as engine pinging, or a seagull cry sound in heart murmurs.

For vision perception (e.g., image recognition) tasks, relatable explanations can be as follows.
1) Saliency can be presented, as is common, as a heatmap to identify important pixels for a decision, such as highlighting the eyes and mouth of a happy face~\cite{jain2018hybrid}), or papillary, sclerotic, solid, and hemorrhagic growth patterns for cancer in a histology image~\cite{rakhlin2018deep}.
2) Counterfactual samples can be based on canonical (prototype) or critical (almost ambiguous) examples~\cite{kim2016examples}, such as feminizing male faces~\cite{xiao2018elegant, he2019attgan} or changing a scene from day to night~\cite{zhu2017unpaired}.
3) Cues can include visual cues, such as depth, motion, color, and contrast~\cite{posner1976visual}.

% TODO
% add about similarity score to different class prototypes as another XAI type?

\section{Conclusion}

We presented the XAI Perceptual Processing Framework to unify a set of contrastive, saliency, counterfactual and cues explanations towards relatable explainable AI. The framework was implemented with RexNet, a modular multi-task deep neural network with multiple explanations, trained to predict vocal emotions. 
From qualitative think-aloud and quantitative controlled studies, we found varying usage and usefulness across the relatable contrastive explanations. 
This work gives insights into providing and evaluating relatable contrastive explainable AI for perception applications, and contributes a new basis towards human-centered XAI.

\begin{acks}
This work was supported in part by the Ministry of Education, Singapore under the grant T2EP20121-0040, and was carried out at the NUS Centre for Research in Privacy Technologies (N-CRiPT) and the NUS Institute for Health Innovation and Technology (iHealthtech).
\end{acks}

\bibliographystyle{ACM-Reference-Format}
\bibliography{main}
\clearpage

\appendix
\onecolumn
\captionsetup[figure]{labelfont={bf},font={small},name={Supplementary Fig.},labelsep=period}
\captionsetup[table]{labelfont={bf},font={small},name={Supplementary Table},labelsep=period}
\setcounter{figure}{0}
\setcounter{table}{0}

\section{Appendix}

\subsection{Vocal cues for different emotions}
\quad
\begin{figure}[ht]
    \centering
    % \hspace*{-0.5cm}
    \includegraphics[width=14.8cm]{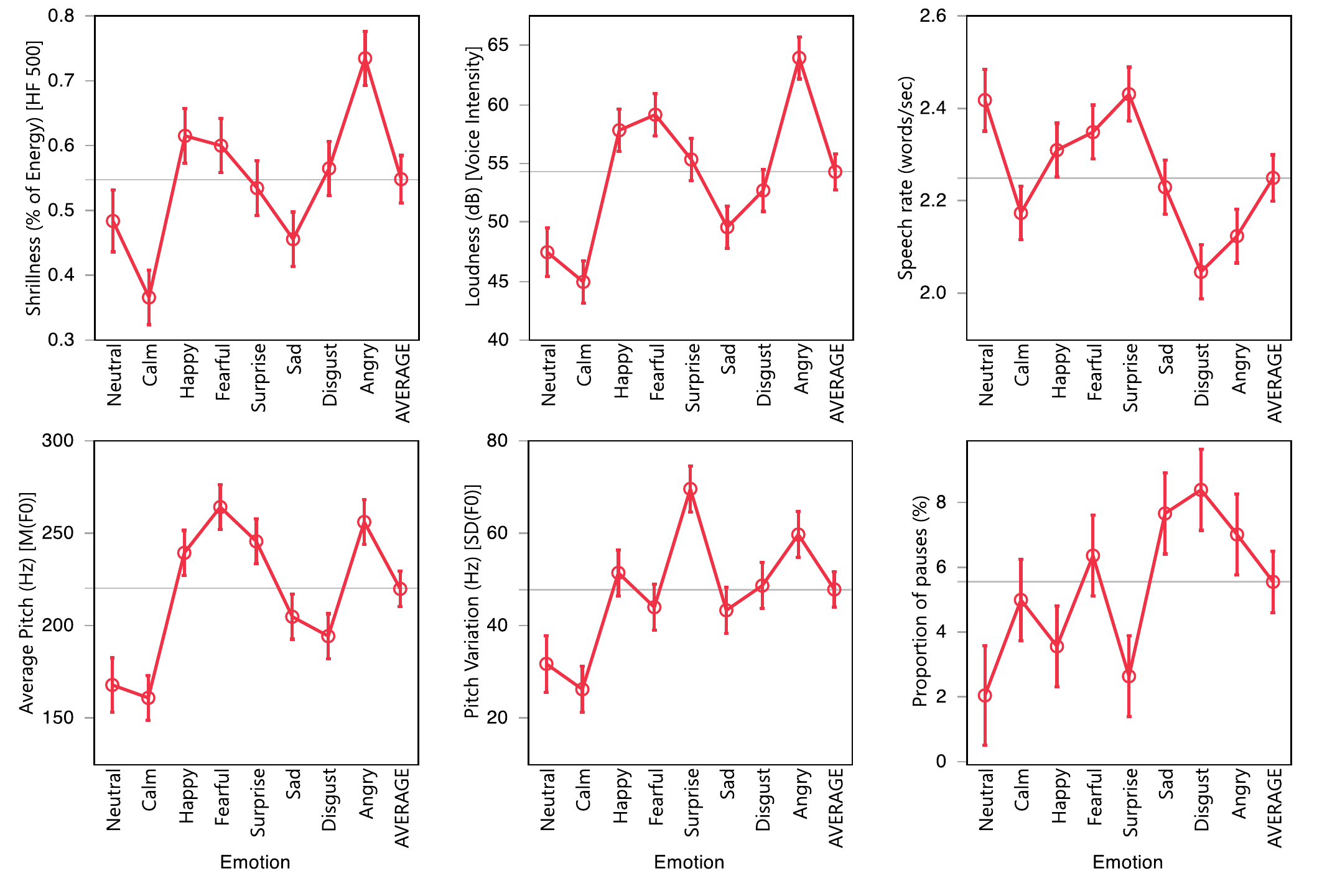}
    \caption{
    Distribution of cue values for different emotions and the average across all voice clips. Values calculated from the RAVDESS dataset~\cite{livingstone2018ryerson}. Differences were used to calculate cue difference relations.
    Grey line indicates average value.
    }
    \label{fig:cue-distros}
    \Description{Line graph of cue value distributions for 8 emotions and the average. 6 cues include Shrillness (\% of Energy), Loudness (dB), Speech rate (words/sec), Average Pitch(Hz), Pitch Variation(Hz) and Proportion of pauses(\%). The grey line indicates the average value.}
    % \vspace{-0.2cm}
\end{figure}
\clearpage

\subsection{User Study Survey}

\begin{figure}[ht]
    \centering    
    \includegraphics[width=12cm]{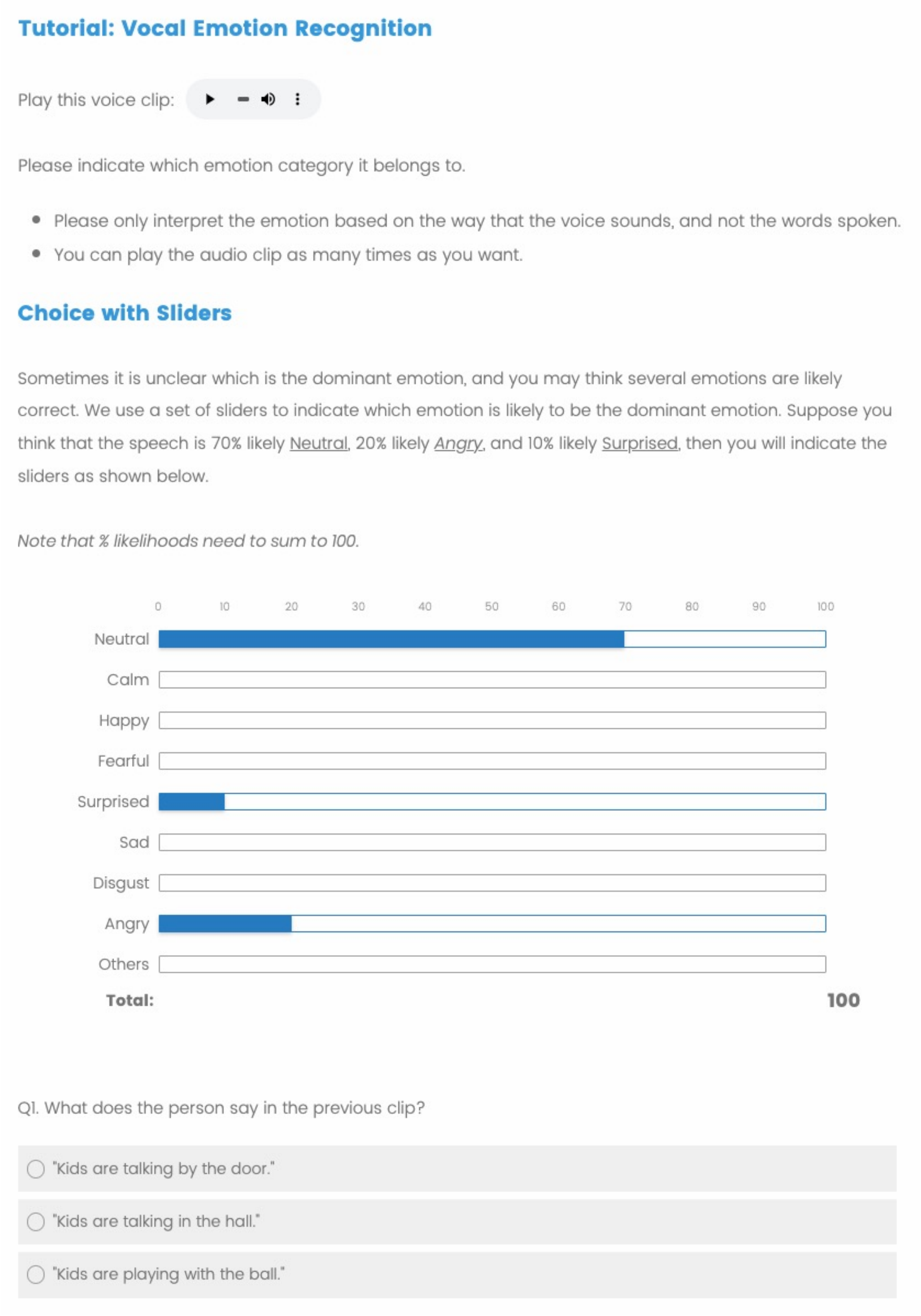}
    \caption{Tutorial to clarify users' tasks, to interpret the “balls and bins” question \cite{goldstein2014lay} and screening question to check users' audio equipment.}
    \label{appendix-figure:slider}
    \Description{Tutorial page of our survey. The page consists of the task description, the instruction on the "balls and bins" question and a screening question.}
\end{figure}

\begin{figure}[ht]
    \centering    
    \includegraphics[width=12cm]{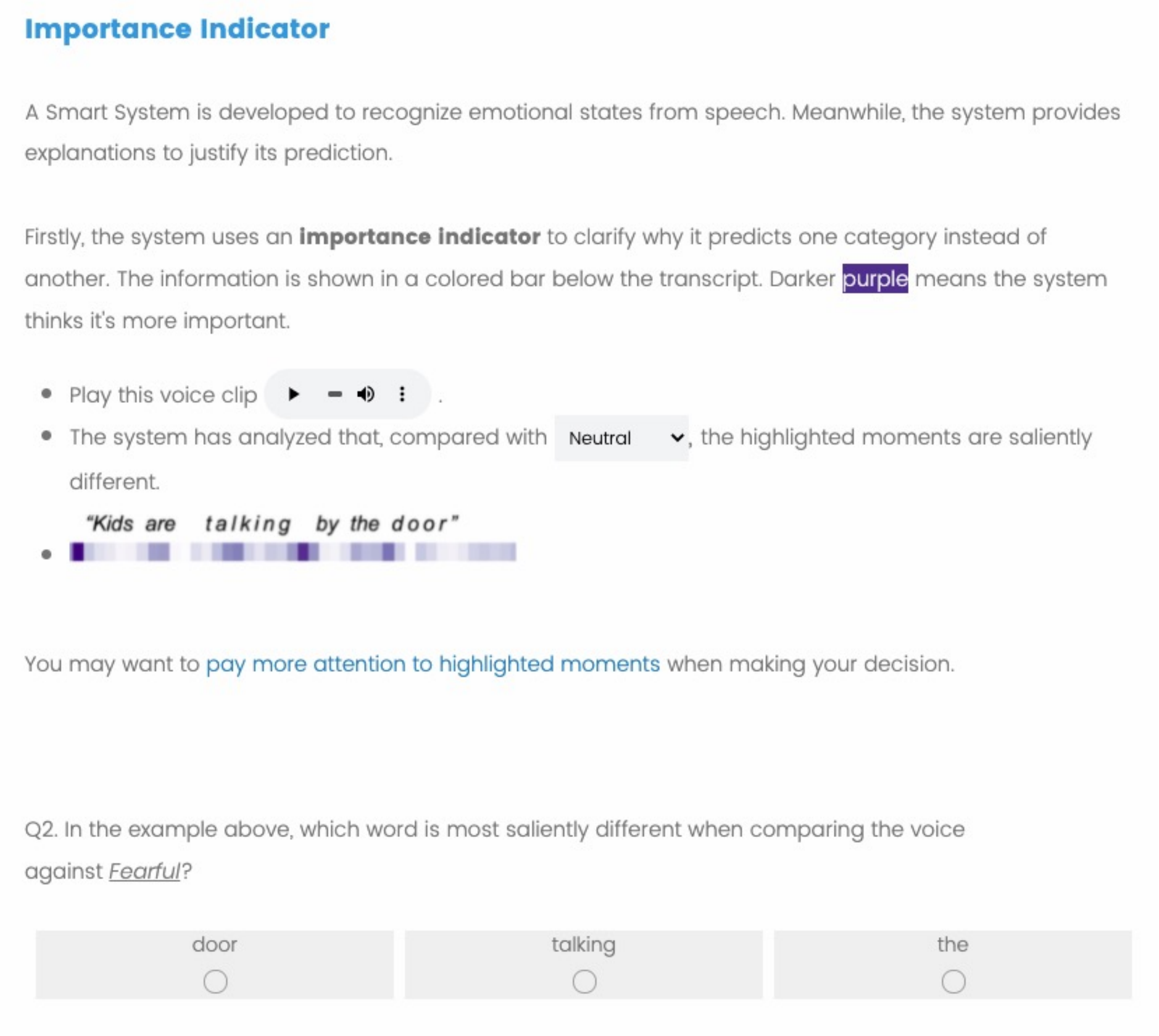}
    \caption{Tutorial on the contrastive saliency explanation and screening question to check users' interpretation.}
    \label{appendix-figure:tutorial_saliency}
    \Description{Tutorial page of our survey. The page consists of the description about the Importance Indicator (contrastive saliency explanation) and a screening question.}
\end{figure}

\begin{figure}[ht]
    \centering    
    \includegraphics[width=12cm]{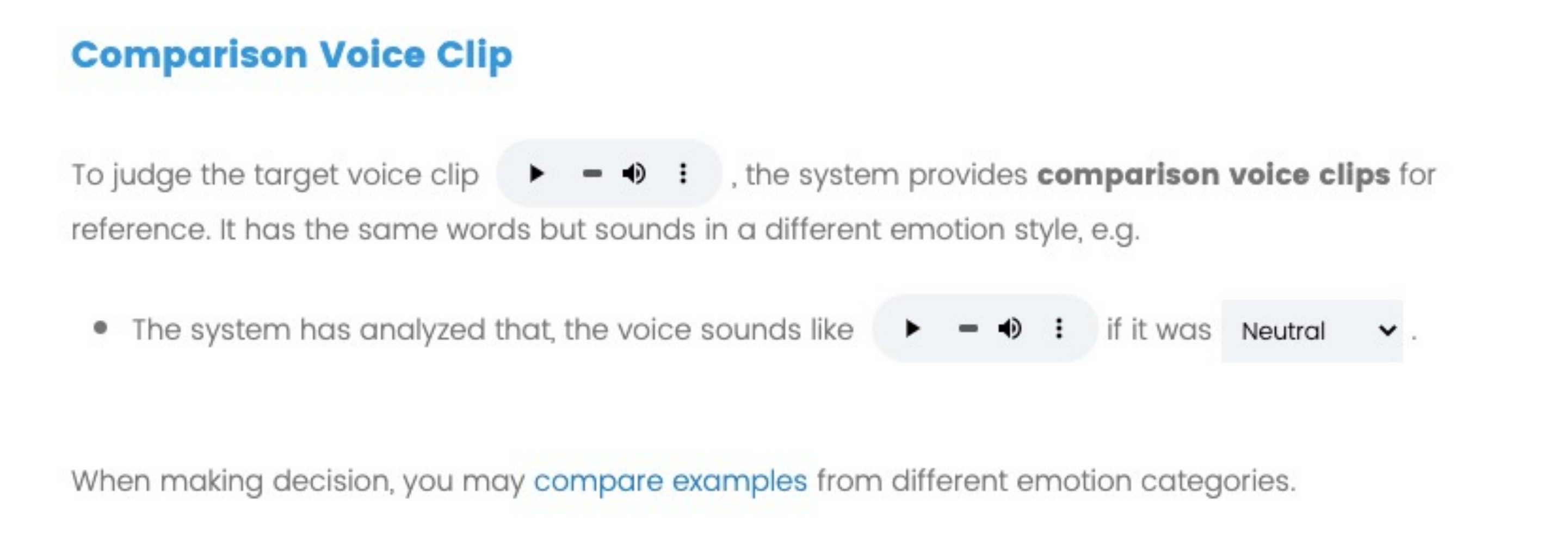}
    \caption{Tutorial on the counterfactual sample explanation.}
    \label{appendix-figure:tutorial_comparisonSample}
    \Description{Tutorial page of our survey. The page consists of the description about the Comparison Voice Clip (counterfactual sample explanation).}
\end{figure}

\begin{figure}[ht]
    \centering    
    \includegraphics[width=13cm]{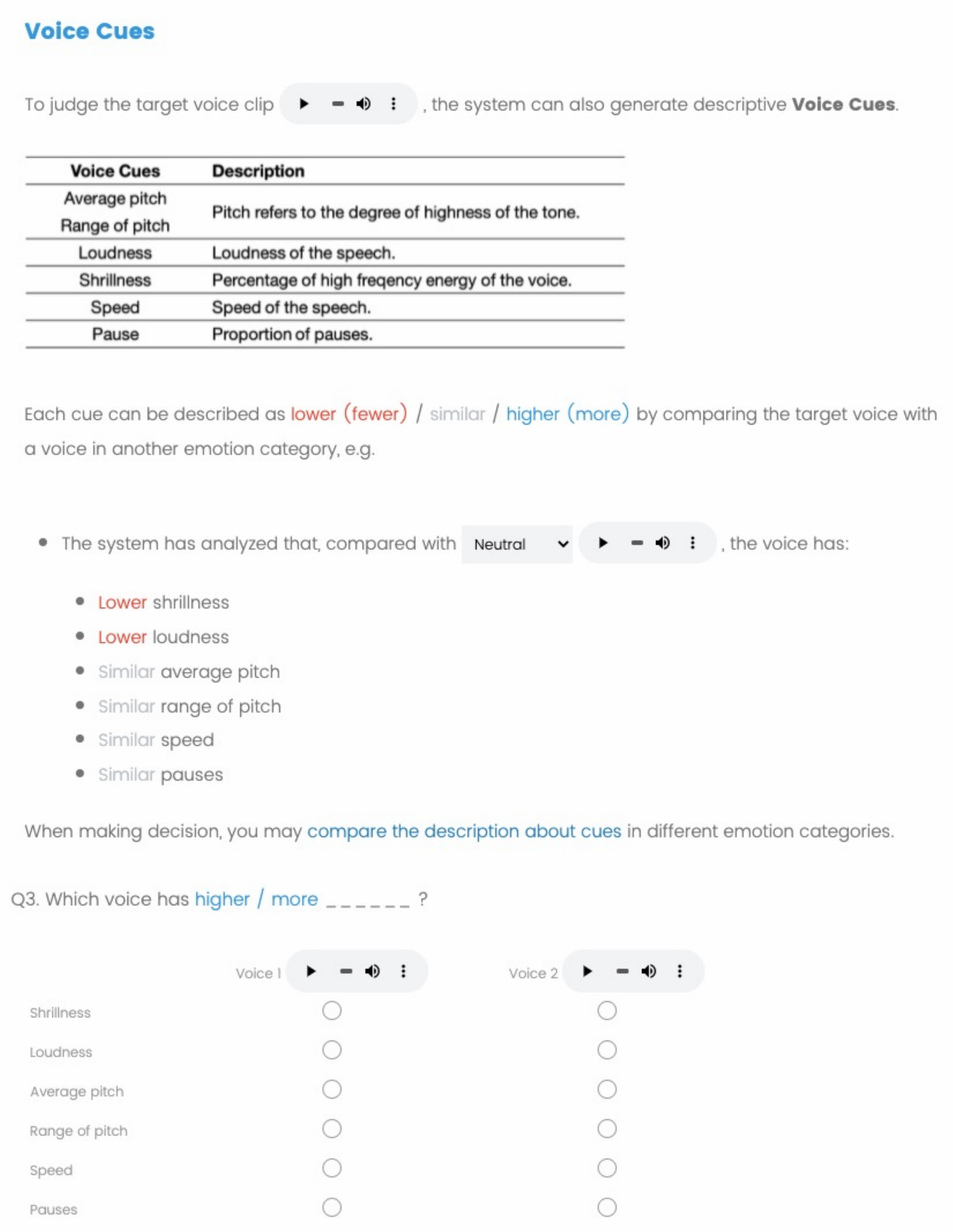}
    \caption{Tutorial on the contrastive cue explanation and screening question to check users' understanding about vocal cues.}
    \label{appendix-figure:tutorial_vocalCues}
    \Description{Tutorial page of our survey. The page consists of the description about the Voice Cues (contrastive cues explanation) and a screening question.}
\end{figure}

\begin{figure}[ht]
    \centering    
    \includegraphics[width=13.5cm]{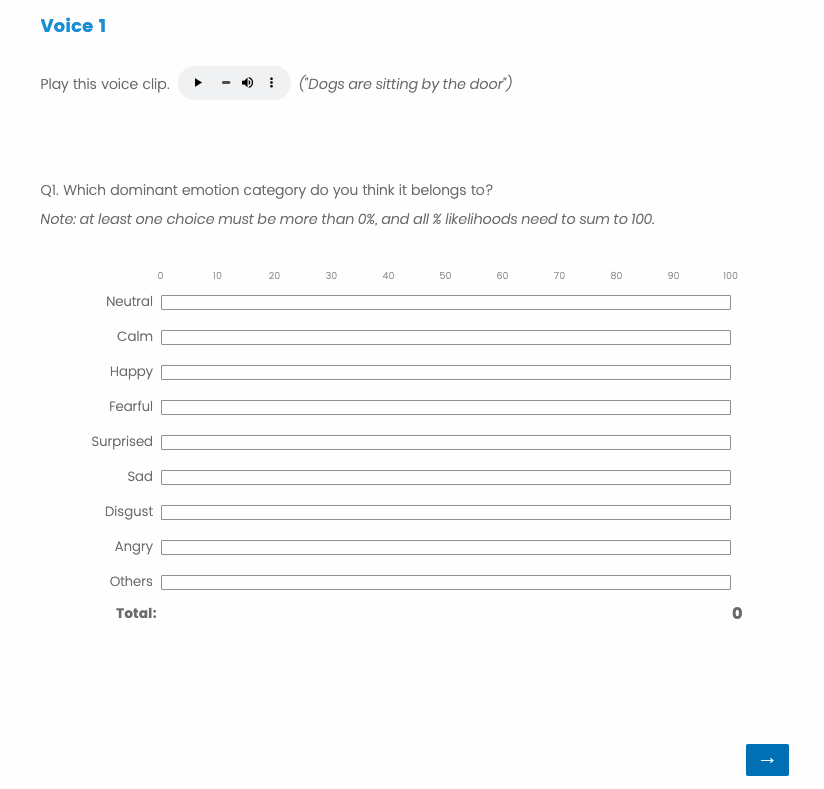}
    \caption{Example practice session per-voice trial before revealing the system's XAI information (Pre-XAI).}
    \label{appendix-figure:trial_prexai}
    \Description{Pre-XAI page in practice session of our survey. The page consists of the audio play button and the "balls and bins" question. }
\end{figure}
\clearpage

\begin{figure}[ht]
    \centering    
    \includegraphics[width=7.5cm]{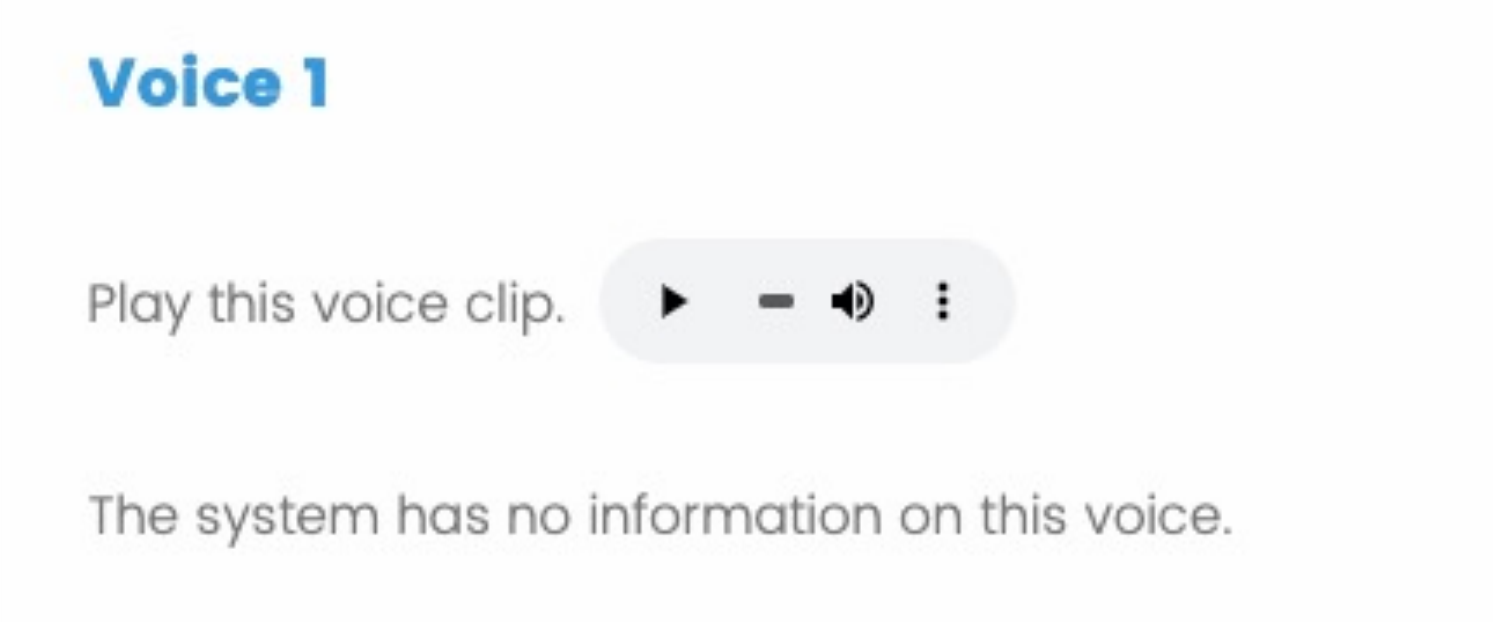}
    \caption{Example main study per-voice trial without the system's explanation.}
    \label{appendix-figure:trial_none}
    \Description{None condition in the task page of our survey. The page consists of the audio play button. }
\end{figure}

\begin{figure}[ht]
    \centering    
    \includegraphics[width=9cm]{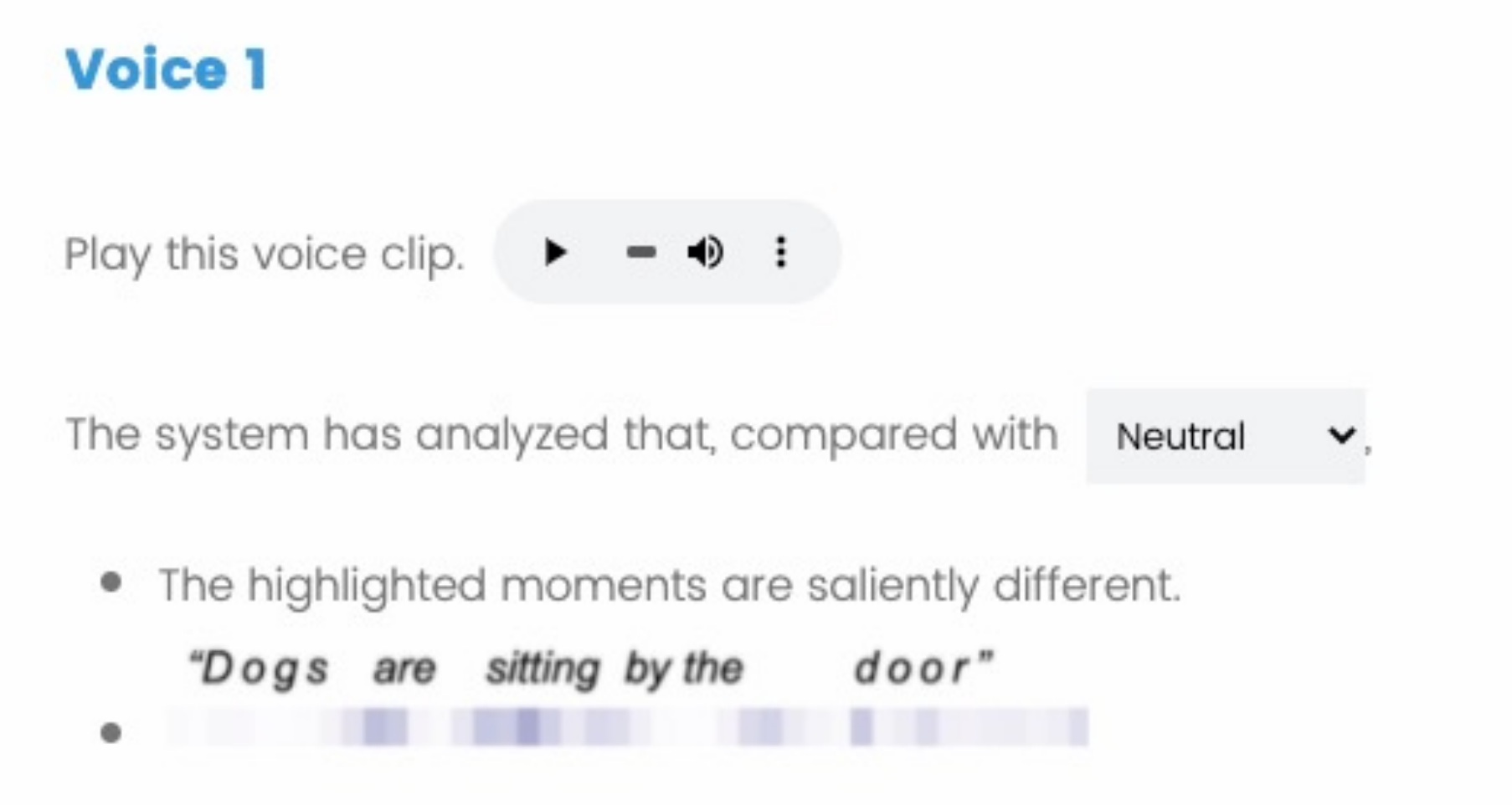}
    \caption{Example main study per-voice trial with the contrastive saliency explanation.}
    \label{appendix-figure:trial_saliency}
    \Description{Saliency condition in the task page of our survey. The page consists of the audio play button and the contrastive saliency explanation. }
\end{figure}

\begin{figure}[ht]
    \centering    
    \includegraphics[width=9.5cm]{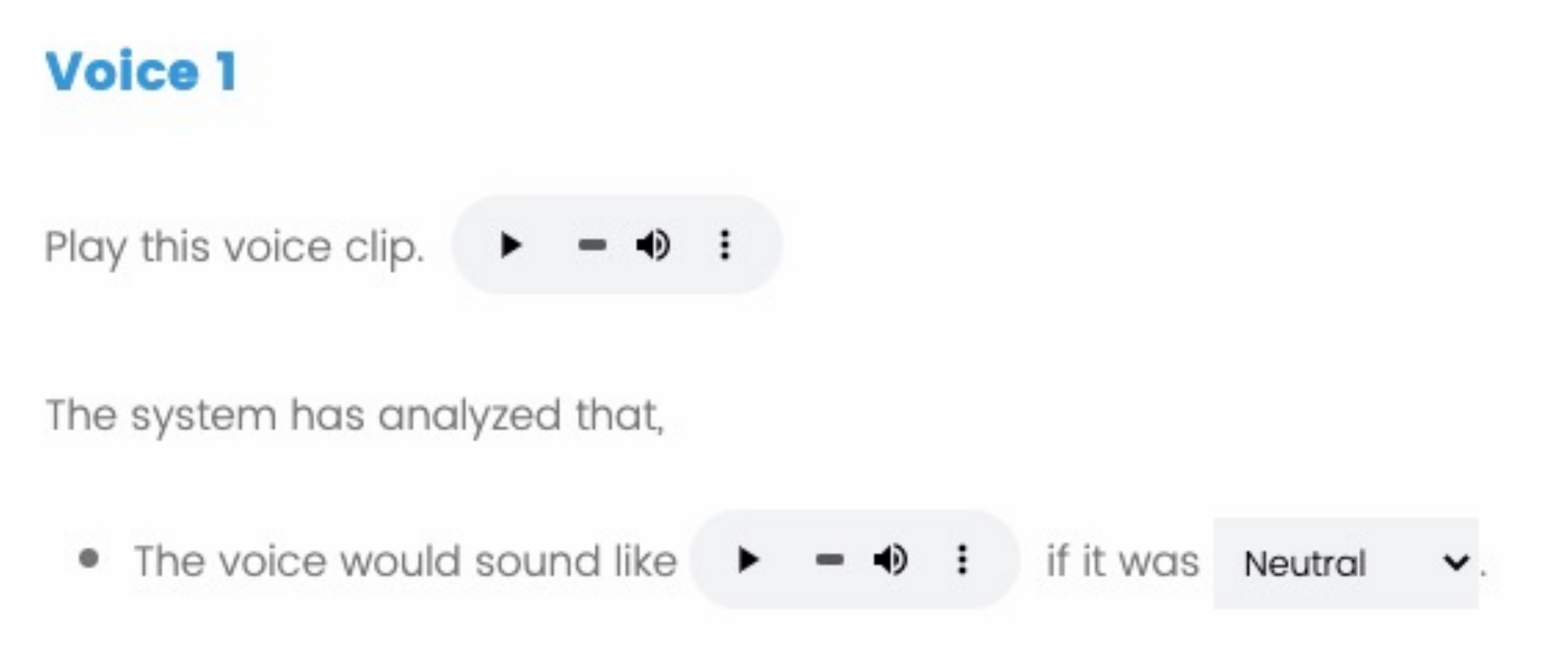} 
    \caption{Example main study per-voice trial with the counterfactual sample explanation.}
    \label{appendix-figure:trial_comparisonVoice}
    \Description{Counterfactual sample condition (C.Sample) in the task page of our survey. The page consists of the audio play button and the counterfactual sample explanation. }
\end{figure}

\begin{figure}[ht]
    \centering    
    \includegraphics[width=10.5cm]{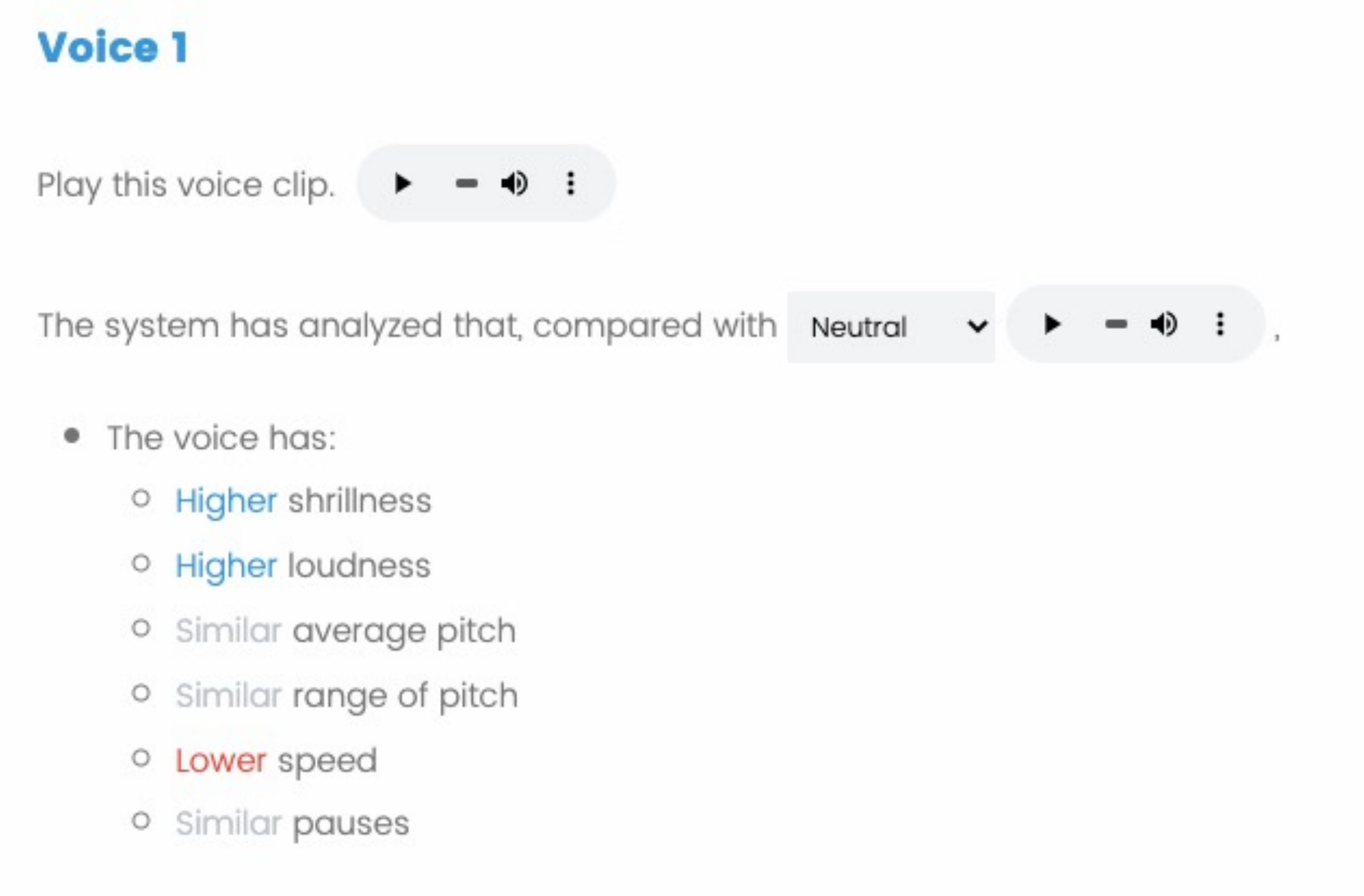}
    \caption{Example main study per-voice trial with the counterfactual sample and contrastive cue explanations.}
    \label{appendix-figure:trial_vocalCue}
    \Description{Counterfactual sample and Contrastive cues condition (C.Sample+Cues) in the task page of our survey. The page consists of the audio play button, the counterfactual sample explanation and contrastive cues explanation. }
\end{figure}

\begin{figure}[ht]
    \centering    
    \includegraphics[width=11.5cm]{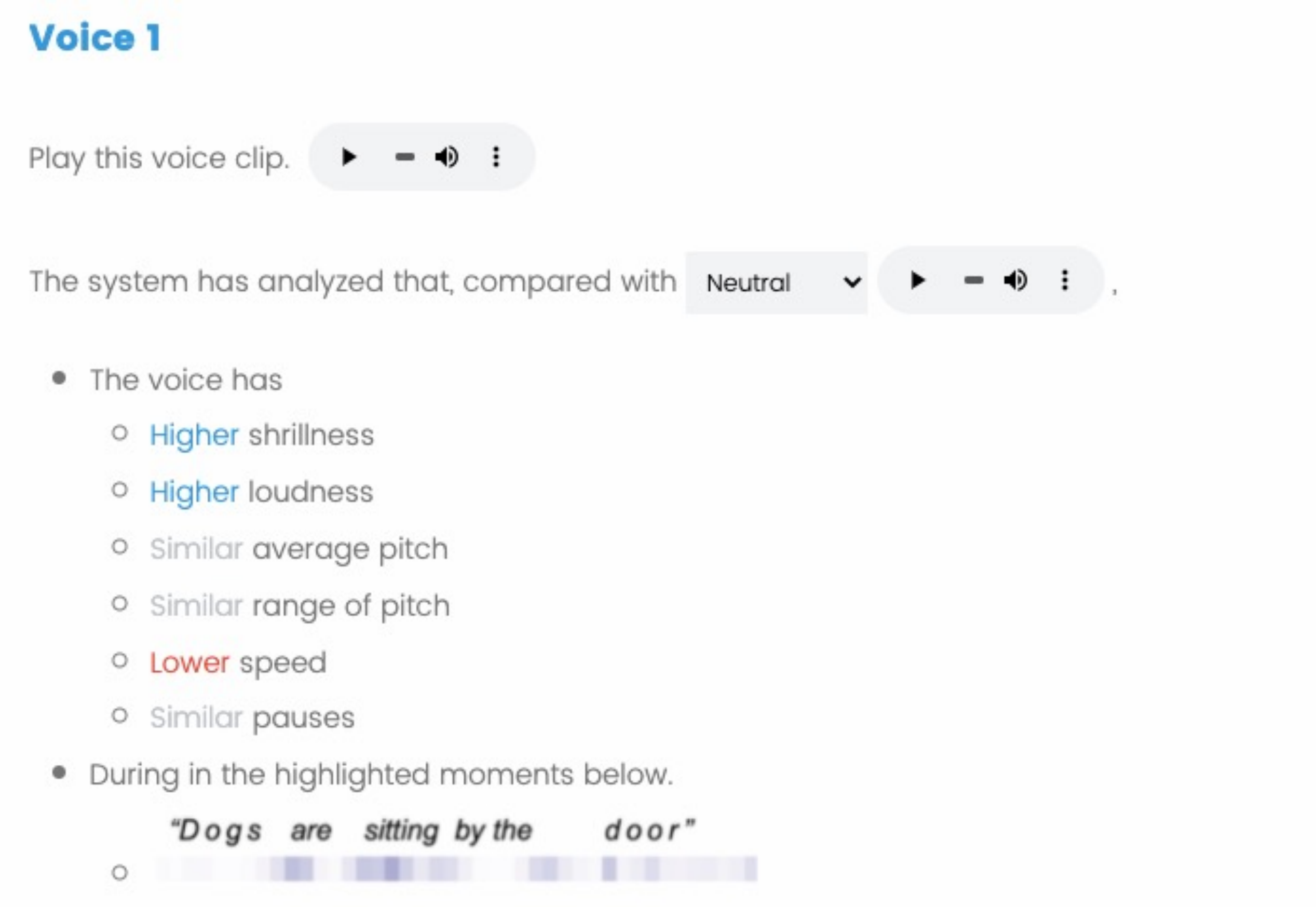}
    \caption{Example main study per-voice trial with the contrastive saliency, counterfactual sample and contrastive cue explanations.}
    \label{appendix-figure:trial_all}
    \Description{All condition (Saliency+C.Sample+Cues) in the task page of our survey. The page consists of the audio play button, the contrastive saliency explanation, the counterfactual sample explanation and contrastive cues explanation. }
\end{figure}

\begin{figure}[ht]
    \centering    
    \includegraphics[width=13cm]{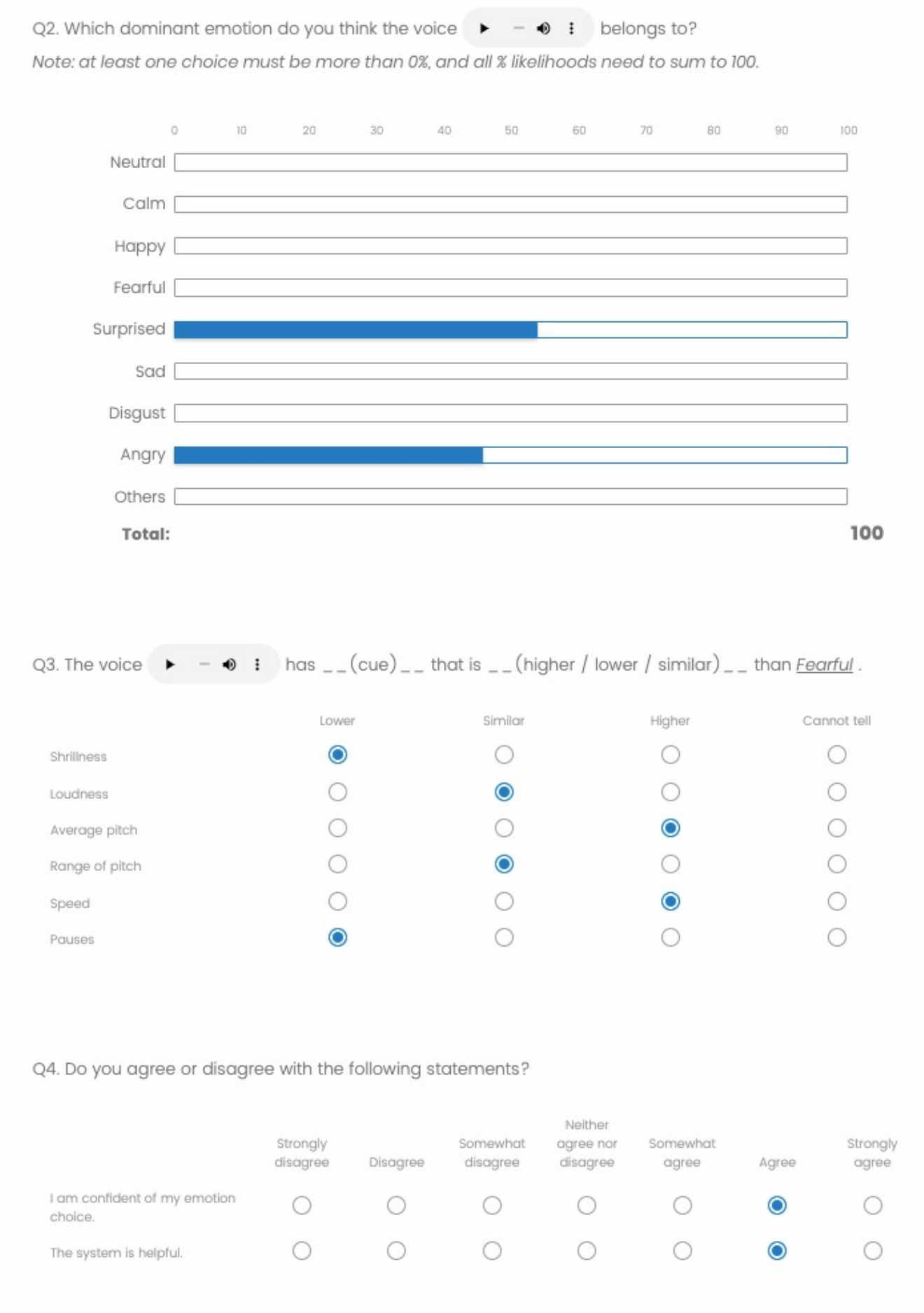}
    \caption{Example main study per-voice trial with the questionnaire after revealing the system's XAI information (Post-XAI).}
    \label{appendix-figure:trial_postxai}
    \Description{Questionnaire in the task page of our survey. The page consists of the audio play button, the "balls and bins" question, the cue relation interpretation question and two Likert-Scale rating questions. }
\end{figure}

\begin{figure}[ht]
    \centering    
    \includegraphics[width=13cm]{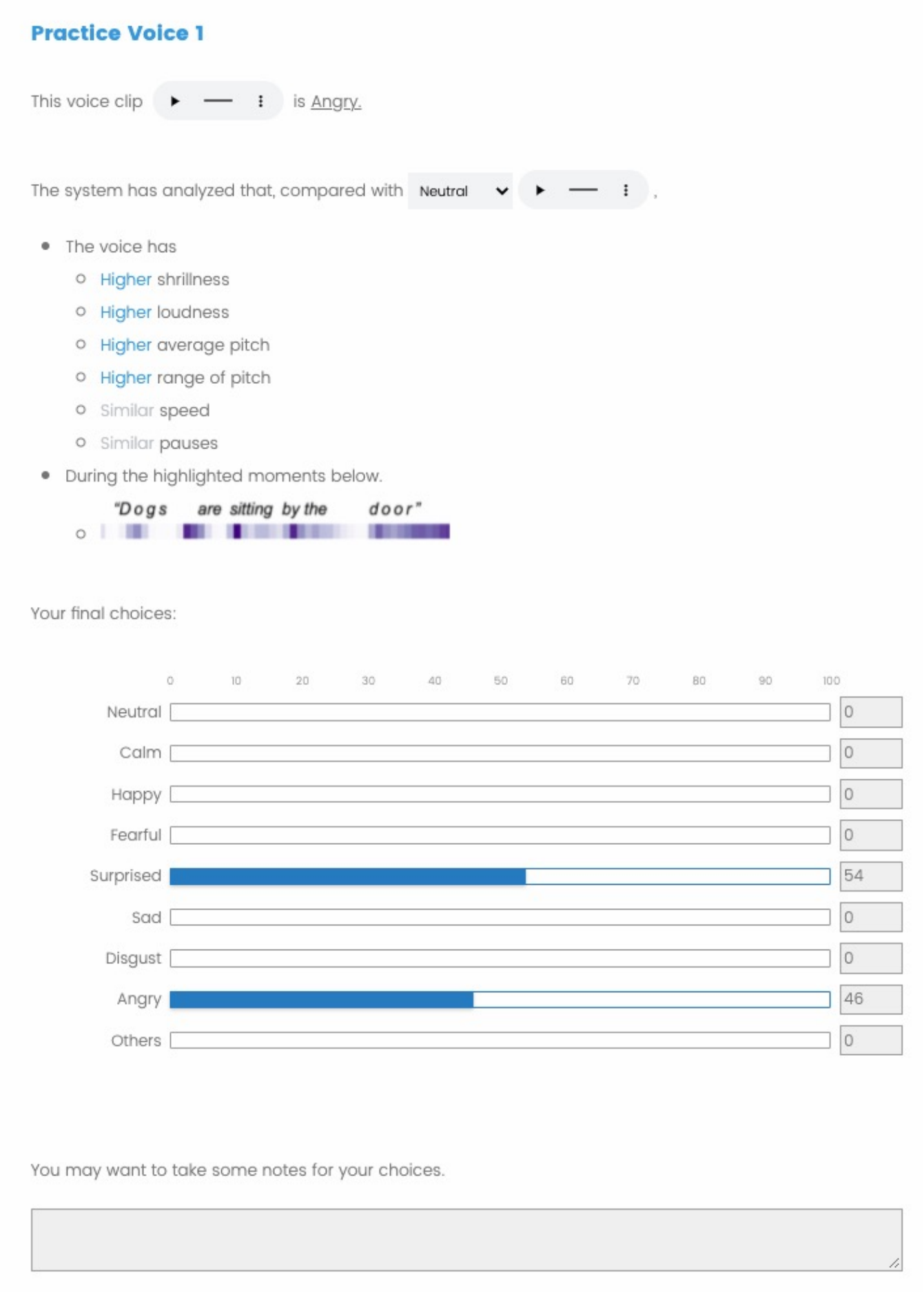}
    \caption{Example practice session per-voice trial to show the correct answer and review users' choices.}
    \label{appendix-figure:trial_review}
    \Description{Review page in practice session of our survey. The page consists of the correct answer of the audio, system's explanations (depending on condition), the user's previous choice and a note box. }
\end{figure}
\clearpage

\subsection{User Study Analysis: Statistical Model}
\begin{table}[ht]
\caption{
% Statistical analysis of responses due to effects as linear mixed eﬀects models.
Statistical analysis of responses due to effects (one per row), as linear mixed effects models with random effects, fixed effects, and their interaction effect. $F$ and $p$ values indicate ANOVA tests and $R^2$ indicate model goodness-of-fit.
}
\begin{tabular}{llrrr}
\hline
Response & \begin{tabular}[c]{@{}l@{}}Linear Effects Model \\ 
(Participants as random effects)\end{tabular} & F     & p\textgreater{}F & $R^2$ \\ 
\hline

\multirow{7}{*}{\begin{tabular}[c]{@{}l@{}}Labeling Correctness\end{tabular}} 
& XAI Type $+$                          & 4.2                      & \textless{}.0030         & .371 \\
& Participant Unaided Skill $+$         & 50.0                     & \textless{}.0001         & \\
& Emotion                               & 68.6                     & \textless{}.0001         & \\
& Voice Clip $+$                        & 67.1                     & \textless{}.0001         & \\
& {\color{lightgray} Trial Number $+$}  & {\color{lightgray} 0.6}  & {\color{lightgray} n.s.} & \\
& Confidence Rating $+$                 & 19.7                     & \textless{}.0001         & \\
& Helpfulness Rating                    & 7.9                      & .0053                    & \\
\arrayrulecolor{lightgray} \hline

\multirow{7}{*}{\begin{tabular}[c]{@{}l@{}}Confidence on Correct Label\end{tabular}} 
& XAI Type $+$                          & 4.2                      & \textless{}.0029         & .435 \\
& Participant Unaided Skill $+$         & 64.4                     & \textless{}.0001         & \\
& Emotion                               & 81.8                     & \textless{}.0001         & \\
& Voice Clip $+$                        & 80.2                     & \textless{}.0001         & \\
& {\color{lightgray} Trial Number $+$}  & {\color{lightgray} 0.3}  & {\color{lightgray} n.s.} & \\
& Confidence Rating $+$                 & 51.9                     & \textless{}.0001         & \\
& Helpfulness Rating                    & 6.4                      & .0113                    & \\
\arrayrulecolor{black} \hline

\multirow{5}{*}{\begin{tabular}[c]{@{}l@{}}Confidence Rating\end{tabular}} 
& {\color{lightgray} XAI Type $+$}                  & {\color{lightgray} 0.5}  & {\color{lightgray} n.s.} & .491 \\
& {\color{lightgray} Participant Unaided Skill} $+$ & {\color{lightgray} 0.1}  & {\color{lightgray} n.s.} & \\
& Emotion                                           & 4.0                      & .0002                    & \\
& Voice Clip $+$                                    & 7.7                      & \textless{}.0001         & \\
& {\color{lightgray} Trial Number}                  & {\color{lightgray} 0.3}  & {\color{lightgray} n.s.} & \\
\arrayrulecolor{lightgray} \hline

\multirow{5}{*}{\begin{tabular}[c]{@{}l@{}}Helpfulness Rating\end{tabular}} 
& XAI Type $+$                          & 5.7                      & .0002                    & .831 \\
& Participant Unaided Skill $+$         & 6.6                      & .0114                    & \\
& Emotion                               & 2.2                      & .0283                    & \\
& Voice Clip $+$                        & 7.7                      & \textless{}.0001         & \\
& {\color{lightgray} Trial Number}      & {\color{lightgray} 0.7}  & {\color{lightgray} n.s.} & \\
\arrayrulecolor{lightgray} \hline

\multirow{5}{*}{\begin{tabular}[c]{@{}l@{}}Log$_{10}$(Task Time)\end{tabular}} 
& {\color{lightgray} XAI Type $+$}                  & {\color{lightgray} 1.0}  & {\color{lightgray} n.s.} & .536 \\
& {\color{lightgray} Participant Unaided Skill} $+$ & {\color{lightgray} 2.8}  & {\color{lightgray} n.s.} & \\
& {\color{lightgray} Emotion} $+$                   & {\color{lightgray} 2.8}  & {\color{lightgray} n.s.} & \\
& Voice Clip $+$                                    & 9.3                      & \textless{}.0001         & \\
& Trial Number                                      & 4.5                      & .0001                    & \\
\arrayrulecolor{black} \hline

\end{tabular}
\label{table:statModelDetails}
\end{table}

\end{document}